\input phyzzx
\newcount\mongocount
\mongocount=1
\def\Figure#1#2#3{
      \vbox to #3in{\hsize=#2in
        \vfil
         \includegraphics{#1}
    }
}
\def\figcap#1#2{
\vtop{\tenpoint\singlespace
\hsize=#1in\smallskip\noindent Figure\ \ \the\mongocount.\ \  #2
\global\advance\mongocount by 1\bigskip}}
\def\mongofigure#1#2#3#4#5{\centerline{\Figure{#1}{#2}{#3}
\figcap{#4}{#5}}}

\hoffset=0.375in
\overfullrule=0pt

\def\dol{{D_{\rm ol}}}
\def\dls{{D_{\rm ls}}}
\def\dos{{D_{\rm os}}}

\def\max{{\rm max}}
\def\min{{\rm min}}
\def\lim{{\rm lim}}

\def\kpc{{\rm kpc}}

\def\kms{{\rm km}\,{\rm s}^{-1}}

\twelvepoint
\font\bigfont=cmr17
\bigskip
\bigskip
\centerline{}
\centerline{}
\centerline{}
\centerline{\bigfont Planet Parameters in Microlensing Events}
\bigskip
\centerline{\bf B.\ Scott Gaudi}
\smallskip
\centerline{and}
\smallskip
\centerline{{\bf Andrew Gould}\footnote{1}{Alfred P.\ Sloan Foundation Fellow}}
\smallskip
\centerline{Dept of Astronomy, Ohio State University, Columbus, OH 43210}
\smallskip
\centerline{e-mail gould@payne.mps.ohio-state.edu, 
gaudi@payne.mps.ohio-state.edu}
\bigskip
\centerline{\bf Abstract}
A planetary microlensing event occurs when a planet perturbs one
of the two images created in a point-mass microlensing event, 
causing a deviation from the standard Paczy\'nski curve.
Determination of the two physical parameters that can be extracted from
a planetary microlensing event, the planet/star mass ratio $q$, 
and the planet/star separation in units of the stellar
Einstein ring, $y_p$, is hampered by several types of degeneracies.  
There are two distinct and qualitatively different classes 
of planetary events: major and minor image perturbations.
For major image perturbations, there is a potentially crippling
continuous degeneracy in $q$ which is of order $\delta_d^{-1}$, where
$\delta_d$ is the maximum fractional deviation of the planetary
perturbation.  Since the threshold of 
detection is expected to be $\delta_d \sim 5\%$, this degeneracy
in $q$ can be a factor of $\sim 
20$.  For minor image perturbations, 
the continuous degeneracy in $q$ is considerably
less severe, and is typically less than a factor $4$.
We show that these degeneracies can be resolved by observations
from dedicated telescopes on several continents together with
optical/infrared photometry from one of these sites.
There also exists a class of discrete degeneracies.
These are typically easy to resolve given good temporal coverage
of the planetary event.  Unambiguous 
interpretation of planetary microlensing events requires the
resolution of both types of degeneracy.  We describe
the degeneracies in detail and specify the situations in
which they are problematic.  
We also describe how individual planet masses and physical
projected separations can be measured.
\bigskip
Subject Headings: gravitational lensing -- planetary systems
\smallskip
\centerline{submitted to {\it The Astrophysical Journal}: 
October 15, 1996}
\centerline{Preprint: OSU-TA-26/96}

\endpage
\normalspace
\chapter{Introduction}

	Two world-wide networks are currently searching for
extra-solar planetary systems by making densely sampled observations of
ongoing microlensing events toward the Galactic bulge (PLANET,
Albrow et al.\ 1996; GMAN, Pratt et al.\ 1996).  
Several other groups will join the search shortly and
there is serious discussion of new initiatives that would intensify the
search by an order of magnitude.  More than 100 microlensing events have
been detected to date by four groups, MACHO (Alcock et al.\ 1996),
EROS (Ansari et al.\ 1996), 
OGLE (Udalski et al.\ 1994), and DUO (Alard 1996) based on observations made
once or twice per night.  The events typically last one week to a few months.
MACHO and OGLE have reported ``alerts'', events detected before peak.  This
alert capability 
is what has allowed PLANET and GMAN to make intensive, sometimes 
round-the-clock, follow-up observations in hopes of finding the planetary
perturbations which are expected to last a day or less.

	In sharp contrast to this explosion of observational activity,
theoretical work on planet detection has been rather sparse, amounting to only
five papers in as many years.  Mao \&
Paczy\'nski (1991) originally suggested that planets might be detected
in microlensing events.  Gould \& Loeb (1992) developed a formalism for
understanding the character of planetary perturbations and made systematic
estimates of the rate of detection for various planetary-system parameters.
Bolatto \& Falco (1994) studied the detection rate in the more general
context of binary systems.  These early works assumed that the lensed star could
be treated as a point source.  The usefulness of this approximation depends
primarily on the angular size of the source $\theta_*$, relative to the
planetary Einstein ring, $\theta_p$,
$$\theta_p = \biggl({m\over M}\biggr)^{1/2}\theta_e,\qquad
\theta_e = \biggl({4 G M \dls\over c^2\dol\dos}\biggr)^{1/2}.\eqn\thetap$$
Here $\theta_e$ is the Einstein ring of the lensing star, $m$ and $M$ are the
masses of the planet and its parent star, and
$\dol$, $\dls$, and $\dos$ are the distances between the observer, lens, and
source.  For Jupiter-mass planets at typical distances $(\dls\sim 2\,\kpc)$
from bulge giant sources, $\theta_p\sim 3\theta_*$ so the approximation is
a reasonable one.  However, for Saturns, Neptunes, and especially Earths,
the finite size of the source becomes quite important, and even for Jupiters
it is not completely negligible.  Moreover, as we will stress below, it is
quite possible to mistake a ``Jupiter event'' in which the source size is
negligible for a ``Neptune event'' with $\theta_*>\theta_p$.  Hence it
is essential to understand finite-source effects even to interpret events
where the source size is in fact small.

	Progress on finite-source effects was substantially delayed by 
problems of computation.  Like all binary lenses, planetary-systems have
caustics, curves in the source plane where a point-source is infinitely 
magnified as two images 
either appear or disappear.  If one attempts to integrate the magnification
of a finite source that crosses a caustic, one is plagued with numerical
instabilities near the caustic.  While it is straight forward to solve these
problems for any given geometry, the broad range of possible geometries makes
it difficult to develop an algorithm sufficiently robust for a statistical
study of lensing events.  Bennett \& Rhie (1996) solved this problem by
integrating in the image plane (where the variation of the magnification is
smooth) rather than the source plane (where it is discontinuous).  They were
thereby able to investigate for the first time the detectability of Earth
to Neptune mass planets.  Gould \& Gaucherel (1996) showed that this approach
could be simplified from a two-dimensional integral over the image of the
source to a one-dimensional integral over its boundary.
The implementation of this method requires some care.  We describe
the practical procedures elsewhere (Gaudi 1996).
The difficult computational problems originally posed by 
finite-source effects are now completely solved.

	To date, the analysis of planetary-system lensing events has focused
on the question of ``detectability'' which was quantified by Gould \& Loeb
(1992) as a certain minimal fractional deviation from a standard
Paczy\'nski (1986) light curve having magnification
$$A(x) = {x^2+ 2\over x(x^2+4)^{1/2}},\qquad x(t) = \biggl[{(t-t_0)^2\over
t_e^2}+\beta^2\biggr]^{1/2},\eqn\aofx$$
where $x$ is the projected lens-source separation in units of $\theta_e$.
Note that this curve is characterized by just three parameters: $t_0$ the
time of closest approach, $\beta$, the impact parameter in units of 
$\theta_e$, and $t_e$, the Einstein radius crossing time.  Bennett \& Rhie
(1996) adopted a similar approach but added the qualification that the
deviation persist for a certain minimum time.

	Here we investigate a different question:  How well can the parameters
of the planetary-system be measured?  As discussed by Gould \& Loeb (1992), 
if there are light-curve data of ``sufficient quality'', two planetary-system 
parameters can generically be extracted from a microlensing event that displays
a planetary perturbation.  These are the planet/star mass ratio, $q$, and the 
planet/star projected separation in units of the stellar Einstein ring, 
$y_p$,
$$q \equiv {m\over M},\qquad y_p \equiv {a_p\over r_e}.\eqn\qdef$$
Here $a_p$ is the physical projected separation, and $r_e=\dol\theta_e$.
As we discuss in \S\ 8,
it will often be possible to make additional observations that specify the
mass and distance of the lensing star, or equivalently $M$ and $r_e$.  
For these cases, the measurements of $q$ and $r_e$ yield
the mass $m=q M$ and projected separation $a_p = y_p r_e$.

If a planet were detected by observing a deviation from the standard curve, but its
mass ratio remained uncertain by a factor of 10, the scientific value of the 
detection would be severely degraded.  Indeed such ``detections''
would probably not receive general acceptance.  Thus, the problems of planet
detection and parameter measurement are intimately connected.
Microlensing planet-detection programs must monitor a total of at least
several hundred events in order to obtain representative statistics on the
frequency of planets.  These observations require large blocks of 1--2 meter
class telescope time coordinated over several continents.  For funding 
agencies and time allocation committees to make rational decisions
about the allocation of scarce resources, and for observers to make rational
choices among prospective targets, it is essential to determine what are the
minimum observational requirement for detecting planetary systems {\it and}
measuring the characteristics of the detected systems.

\chapter{Types of Degeneracy}
\section{Discrete}
\FIG\one{
Discrete degeneracies.  Panel (a) shows a lensing light curve
with ({\it solid curve}) and without ({\it dashed curve}) taking account of
the presence of a planet with mass ratio $q=10^{-3}$.  
Panel (b) shows the associated lensing geometry.
The two solid curves represent the path of the images relative to the lens.
The crosses represent the image positions 
at the time of the perturbation.  
The circles are the four planet positions for which the light curves
reproduce the measured parameters $\delta_d$ (maximum fraction deviation)
and $t_d$ (FWHM of deviation) at the peak of the disturbance when
the source-lens separation is $x_d$.
The filled circle is the ``actual'' planet position.
Panel (c) shows the four associated light curves for times
near the peak of the perturbation, $t_{0,d}$.  Note that time
is expressed in units of the perturbation time scale, $t_d$, not
$t_e$.  The bold curve
corresponds to the ``actual'' planet position.  Clearly, if the light curve
is well sampled, the two dashed curves corresponding to the image position 
inside the Einstein ring in panel (b) could be ruled out immediately.  
However, the two solid curves are less easily distinguished. These
differ by $\sim 15\%$ in planet/star separation and $10\%$ in mass.
See panel (b) and Table 1.
}

\topinsert
\mongofigure{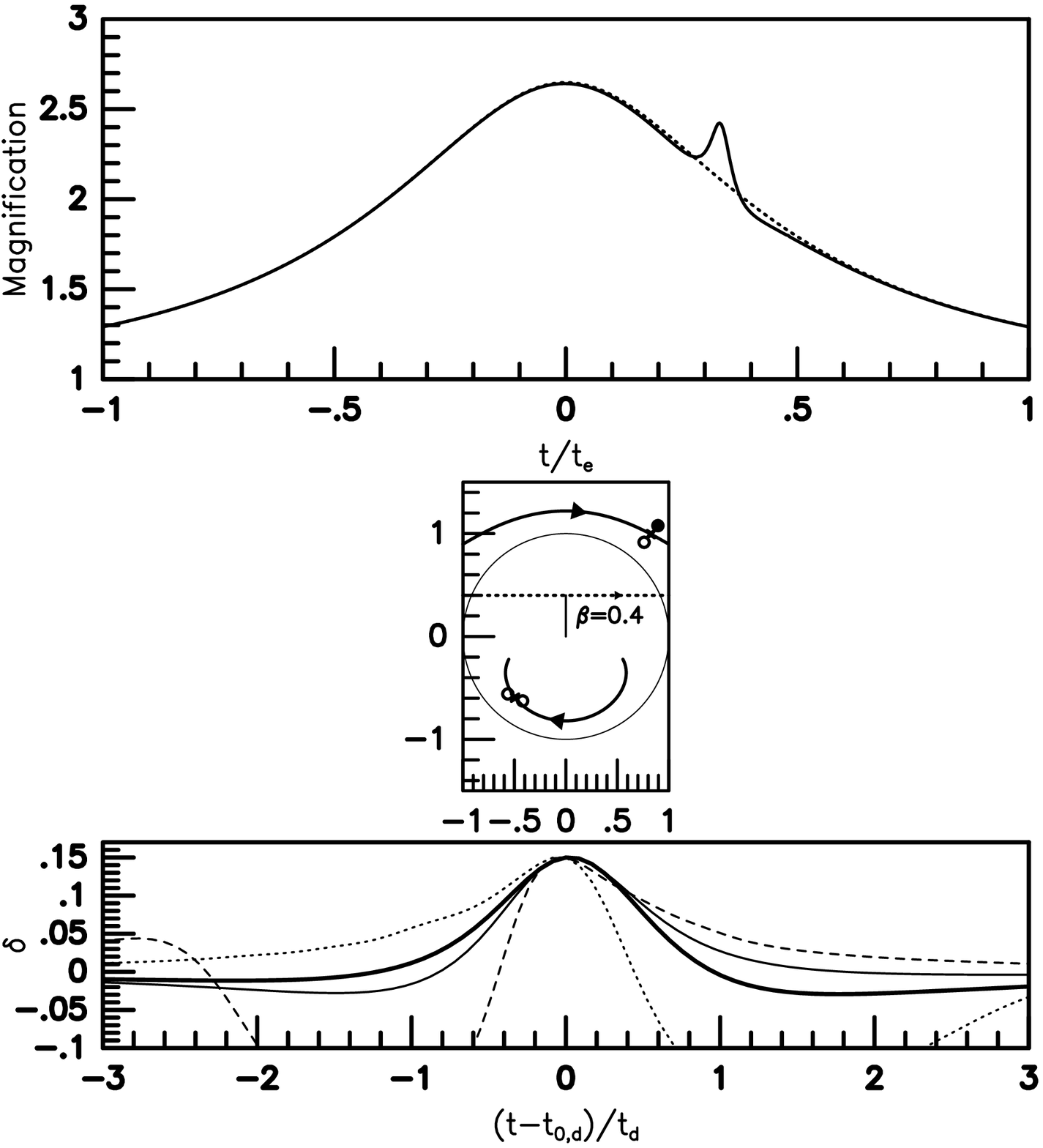}{6.4}{6.5}{7.0}
{
Discrete degeneracies.  Panel (a) shows a lensing light curve
with ({\it solid curve}) and without ({\it dashed curve}) taking account of
the presence of a planet with mass ratio $q=10^{-3}$.  
Panel (b) shows the associated lensing geometry.
The two solid curves represent the path of the images relative to the lens.
The crosses represent the image positions 
at the time of the perturbation.  
The circles are the four planet positions for which the light curves
reproduce the measured parameters $\delta_d$ (maximum fraction deviation)
and $t_d$ (FWHM of deviation) at the peak of the disturbance when
the source-lens separation is $x_d$.
The filled circle is the ``actual'' planet position.
Panel (c) shows the four associated light curves for times
near the peak of the perturbation, $t_{0,d}$.  Note that time
is expressed in units of the perturbation time scale, $t_d$, not
$t_e$.  The bold curve
corresponds to the ``actual'' planet position.  Clearly, if the light curve
is well sampled, the two dashed curves corresponding to the image position 
inside the Einstein ring in panel (b) could be ruled out immediately.  
However, the two solid curves are less easily distinguished. These
differ by $\sim 15\%$ in planet/star separation and $10\%$ in mass.
See panel (b) and Table 1.
}
\endinsert

	Planetary-system lensing events are subject to two different
discrete degeneracies.  The first ambiguity relates to which image the
planet is perturbing: the major image outside the Einstein ring or the minor
image inside the Einstein ring.  For almost all cases, this degeneracy is
easily broken provided there is good temporal coverage of the light curve.
However, if it is not broken the uncertainty in $q$ and $y_p$ can be a factor
of a few.  The magnitudes of these uncertainties depend only on the overall
geometry of the event and not on the mass of the planet.  The second ambiguity
relates to whether the planet lies closer to or farther from the star than
does the position of the source image that it is perturbing.  This degeneracy
is more difficult to break,  but it does not seriously affect the determination
of $q$, and the uncertainty induced in $y_p$ is proportional to 
$q^{1/2}$ and is therefore often much smaller than the one induced by first
degeneracy.  These two discrete degeneracies are illustrated in Figure \one.
The values of $q$ and $y_p$ for each of the four possible solutions are 
displayed in Table 1.  

$$\vbox{\halign{#\hfil\quad&\hfil#\quad&\hfil#
\quad&\hfil#\hfil\quad&\hfil#\hfil\quad&\hfil#\hfil\cr
\multispan{3}{\hfil TABLE 1 \hfil}\cr
\noalign{\medskip}
\multispan{3}{\hfil Degenerate Parameter Values: Discrete\hfil}\cr
\noalign{\smallskip}
\noalign{\hrule}
\noalign{\smallskip}
\noalign{\hrule}
\noalign{\smallskip}
&\hfil planet/star \hfil&\hfil planet/star
\hfil\cr
&\hfil separation\hfil&\hfil mass ratio
\hfil\cr
&\hfil $y_p$ \hfil&\hfil $q/q_0$ \hfil\cr
\noalign{\smallskip}
\noalign{\hrule}
\noalign{\smallskip}
Major Image&\hfil1.40\hfil&\hfil1.00\hfil \cr
&\hfil1.19\hfil&\hfil0.91\hfil \cr
Minor Image&\hfil0.75\hfil&\hfil1.08\hfil \cr
&\hfil0.80\hfil &\hfil0.88\hfil \cr
\noalign{\smallskip}
\noalign{\hrule}
}}
$$

\FIG\two{
Ten light curves with shear $\gamma=0.6$, $\phi=90^\circ$
(see eqs.\ 3.1 and 3.2), all with maximum 
deviation $\delta_d=10\%$ and FWHM
$t_d = 0.06\,t_e$.  The ratios of source radius to planet Einstein ring range
from $\rho=0.1$ to $\rho=2.87$, the largest source radius consistent
with this maximum deviation.  Table 2 gives the corresponding
values of $q=m/M$, and
proper motion, $\mu$, relative to the fiducial values $q_0$ and
$\mu_0$ at the arbitrarily chosen value $\rho=0.3$.
}

\topinsert
\mongofigure{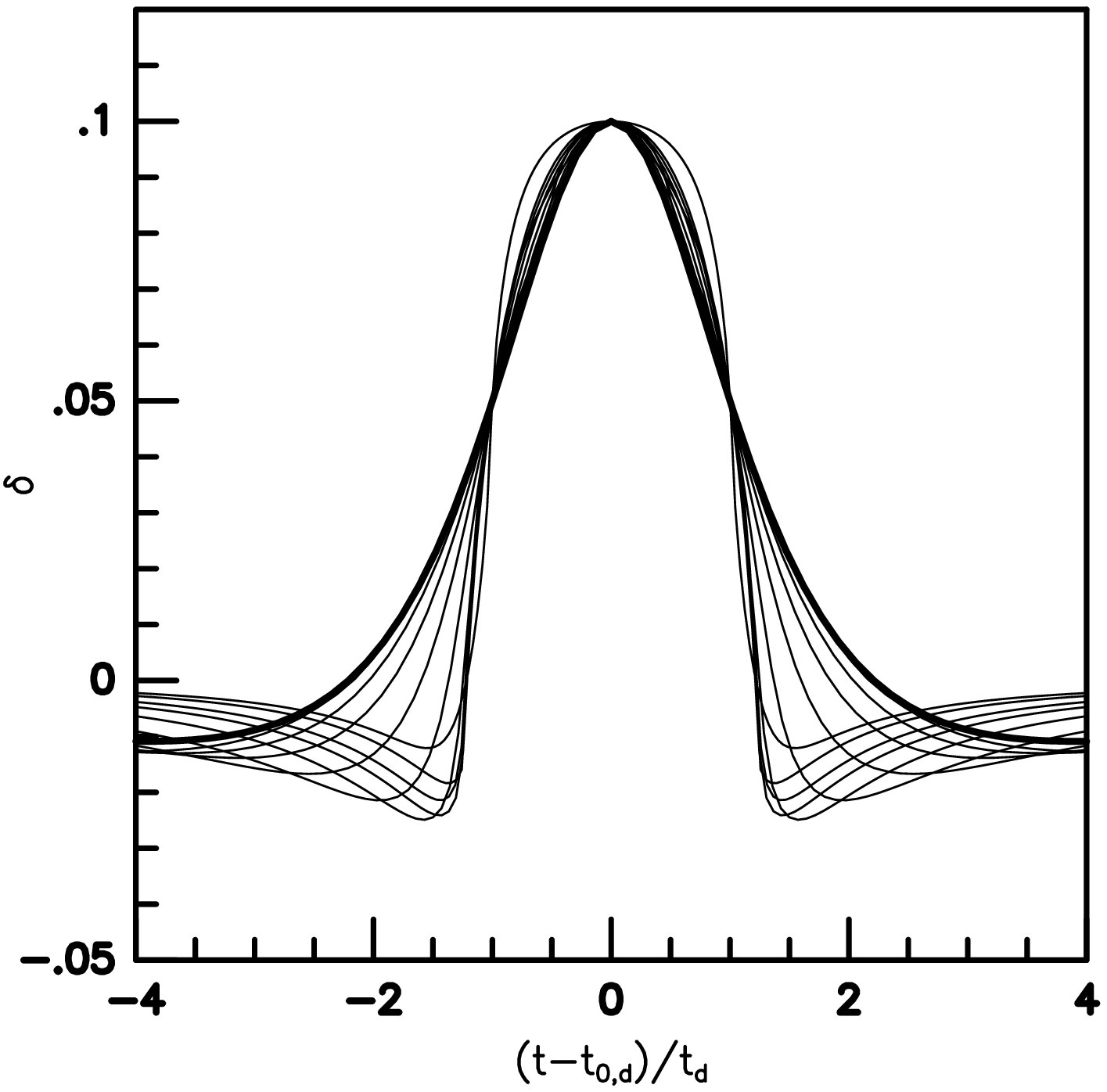}{6.4}{5.5}{6.0}
{
Ten light curves with shear $\gamma=0.6$, $\phi=90^\circ$
(see eqs.\ 3.1 and 3.2), all with maximum 
deviation $\delta_d=10\%$ and FWHM
$t_d = 0.06\,t_e$.  The ratios of source radius to planet Einstein ring range
from $\rho=0.1$ to $\rho=2.87$, the largest source radius consistent
with this maximum deviation.  Table 2 gives the corresponding
values of $q=m/M$, and
proper motion, $\mu$, relative to the fiducial values $q_0$ and
$\mu_0$ at the arbitrarily chosen value $\rho=0.3$.
}
\endinsert

\section{Continuous}

	In addition, there is a continuous 
degeneracy arising from finite-source
effects being misinterpreted as a larger value of $q$.  This is because
$q$ is determined from the (square of the) duration of the planetary 
perturbation relative to the total duration of the event.  If the size of the 
source
is larger than the Einstein ring of the planet, then the duration of the
planetary perturbation will be the crossing time of the source, not of the
planet Einstein ring.  Figure \two\ shows 10 light curves all with the same
maximum fractional deviation, $\delta_d$, 
and same full width half maximum (FWHM) of perturbation, $t_d$.
The parameter that differs in each of these curves is the ratio 
of source radius, $\theta_*$, to planet Einstein radius, $\theta_p=q\theta_e$,
$$\rho = {\theta_*\over \theta_p}.\eqn\rhodef$$
Table 2 gives the inferred values of $q$ 
and of the proper motion $\mu$ (of the planetary
system relative to the observer-source line of sight)
associated with each curve in units of the ``fiducial'' values 
associated with $\rho=0.3$.  In so far as one could not distinguish among these
curves, any of these parameter combinations would be acceptable.  
The fiducial parameters $q_0$ and $\mu_0$ would then be measurable but the
actual values of $\mu$ and $q$ would not.  The proper motion of 
both bulge and disk lenses is typically $\mu \sim {\cal{O}}(V_{\rm{LSR}}/ R_0)
\sim 30\,\kms \kpc^{-1}$, where $V_{\rm{LSR}}\sim 220\, \kms$ is the
rotation speed of the Local Standard of Rest, and $R_0 \sim 8\, \kpc$
is the Galactocentric distance.
If, for the example shown in Table 2, the fiducial value were
measured as $\mu_0 \sim V_{\rm{LSR}}/R_0$, 
one might then choose to argue that the proper motions 
associated with the low-mass solutions (i.e.\ $\mu \sim \mu_0/3$) 
would be so low as to be {\it a priori} unlikely.  
However, these solutions could not actually be 
ruled out by such an argument, since the distribution of
$\mu$ is rather broad (see Han \& Gould 1995).  
Thus, there would remain a factor $\sim 15$
uncertainty in the planet/star mass ratio.

$$\vbox{\halign{#\hfil\quad&\hfil#\quad&\hfil#
\quad&\hfil#\hfil\quad&\hfil#\hfil\quad&\hfil#\hfil\cr
\multispan{3}{\hfil TABLE 2 \hfil}\cr
\noalign{\medskip}
\multispan{3}{\hfil Degenerate Parameter Values: 
Continuous Major Image\hfil}\cr
\noalign{\smallskip}
\noalign{\hrule}
\noalign{\smallskip}
\noalign{\hrule}
\noalign{\smallskip}
\hfil dimensionless \hfil&\hfil 
planet/star \hfil&\hfil proper motion \hfil\cr
\hfil source radius\hfil&\hfil 
mass ratio\hfil&\hfil \hfil\cr
\hfil $\rho$ \hfil&\hfil $q/q_0$ \hfil&\hfil $\mu/\mu_0$ \hfil\cr
\noalign{\smallskip}
\noalign{\hrule}
\noalign{\smallskip}
\hfil0.10\hfil&\hfil1.095\hfil&\hfil2.867\hfil \cr
\hfil0.20\hfil&\hfil1.041\hfil&\hfil1.470\hfil \cr
\hfil0.30\hfil&\hfil1.000\hfil&\hfil1.000\hfil \cr
\hfil0.60\hfil&\hfil0.957\hfil&\hfil0.511\hfil \cr
\hfil0.90\hfil&\hfil0.767\hfil&\hfil0.381\hfil \cr
\hfil1.20\hfil&\hfil0.566\hfil&\hfil0.332\hfil \cr
\hfil1.50\hfil&\hfil0.373\hfil&\hfil0.327\hfil \cr
\hfil1.80\hfil&\hfil0.236\hfil&\hfil0.343\hfil \cr
\hfil2.10\hfil&\hfil0.163\hfil&\hfil0.354\hfil \cr
\hfil2.40\hfil&\hfil0.127\hfil&\hfil0.351\hfil \cr
\hfil2.70\hfil&\hfil0.093\hfil&\hfil0.364\hfil \cr
\hfil2.87\hfil&\hfil0.074\hfil&\hfil0.383\hfil \cr
\noalign{\smallskip}
\noalign{\hrule}
}}
$$

\section{Relation Between Degeneracies in $q$ and $\mu$}

	From the relation $\mu=\theta_e/t_e$, we obtain the 
identity $\mu =(\theta_e/\theta_p)(\theta_p/\theta_*)(\theta_*/t_e)$ or
$$\mu\rho q^{1/2}={\theta_*\over t_e}.\qquad \eqn\murhodeg$$
Since the quantities on the right hand side of this equation are observables,
the product on the left hand side must be constant for all allowed parameter
combinations in any given planetary event: $\mu\rho q^{1/2}=$constant.  This
equation then establishes a relationship between degeneracies in $q$ and
degeneracies in $\mu$.  If a range of solutions are permitted that have
different values of $\rho$ but very similar values of $q$,
then we say that the mass ratio is not degenerate.  However, it follows from 
equation \murhodeg\ that the proper motion $\mu$ then varies inversely as $\rho$ and
therefore that it is degenerate.  Similarly, if the range of allowed solutions
all have the same value of $\mu$, then the proper motion is not degenerate
but then $q\propto \rho^{-2}$ and so the mass ratio is degenerate.
This relationship is illustrated by Table 2.  The region $\rho \le 0.3$
has well-determined $q$ but degenerate $\mu$, while the region
$\rho \gsim 1$ has well-determined $\mu$ but degenerate $q$.

\chapter{The Chang-Refsdal Lens Approximation}

	In order to systematically investigate the role of these degeneracies
and to determine the data that are required to break them, we follow Gould \&
Loeb (1992) and approximate the planetary perturbation as a Chang-Refsdal
lens (Chang \& Refsdal 1979; Schneider, Ehlers, \& Falco 1992).  
A Chang-Refsdal lens is a point mass (in this case the planet) superimposed
on a uniform background shear $\gamma$.  For any given lensing event, the
value of $\gamma$ is simply the shear due to the lensing star at the 
{\it unperturbed} position of the image that is perturbed by the planet.
The evaluation of $\gamma$ is made at the mid-point of the 
perturbation.  The source
position at this midpoint, $x_d$ is known from the light curve 
(see Figs.\ \one-a,b).  The associated image positions are $y_{d,\pm}$ with
shears $\gamma_\pm$ given by
$$\gamma_\pm = y_{d,\pm}^{-2},\qquad y_{d,\pm} 
={(x_d^2+4)^{1/2}\pm x_d\over 2}.
\eqn\gammadef$$
Thus the shear is known (up to a two-fold ambiguity) simply from the position
of the planetary perturbation on the overall light curve.

\FIG\three{
Chang-Refsdal magnification contours of a point source as
a function of source position in units of the planet Einstein ring, $\theta_p=q\theta_e$,
for various pairs of shears
$(\gamma_+$ and $\gamma_-=\gamma_+^{-1})$ corresponding to planetary 
perturbations 
of the major and minor images, respectively. 
Magnification contours are calculated including the contribution
of the unperturbed image. Contour pairs are for
$\gamma_+=0.8,$ 0.6, 0.4, and 0.2 in panels (a,b), (c,d), (e,f), and (g,h),
and correspond to source positions at perturbation of 
$x_d = 0.22,$ 0.52, 0.95, and 1.79.  Super-bold contour is no deviation.
Bold contours are $\delta = 5\%$, 10\%, 20\%, and $\infty$.  Non-bold contours
are $-5\%$,  $-10\%$,  and $-20\%$.  Diagonal lines in panels (c) and (d)
represent possible trajectories assuming that the overall light curve shows
$x_d=0.52$ (i.e.\ $\gamma_+=0.6$) and $\beta=0.4$ (i.e.\ $\phi=\sin^{-1}{\beta/
x_d} \sim 50^\circ$).
If the maximum deviation were observed to be $\delta_d=20\%$ (and the 
point-source approximation were known to be valid), then the trajectory
must be either B or D. 
}

When computing the fractional deviation, $\delta$, of the
Chang-Refsdal lens from the standard Paczy\'nski curve, one always
normalizes relative to the total unperturbed magnification [eq.\ \aofx]
which includes both the image perturbed by the planet and the image that
remains unperturbed (Gould \& Loeb 1992).

\topinsert
\mongofigure{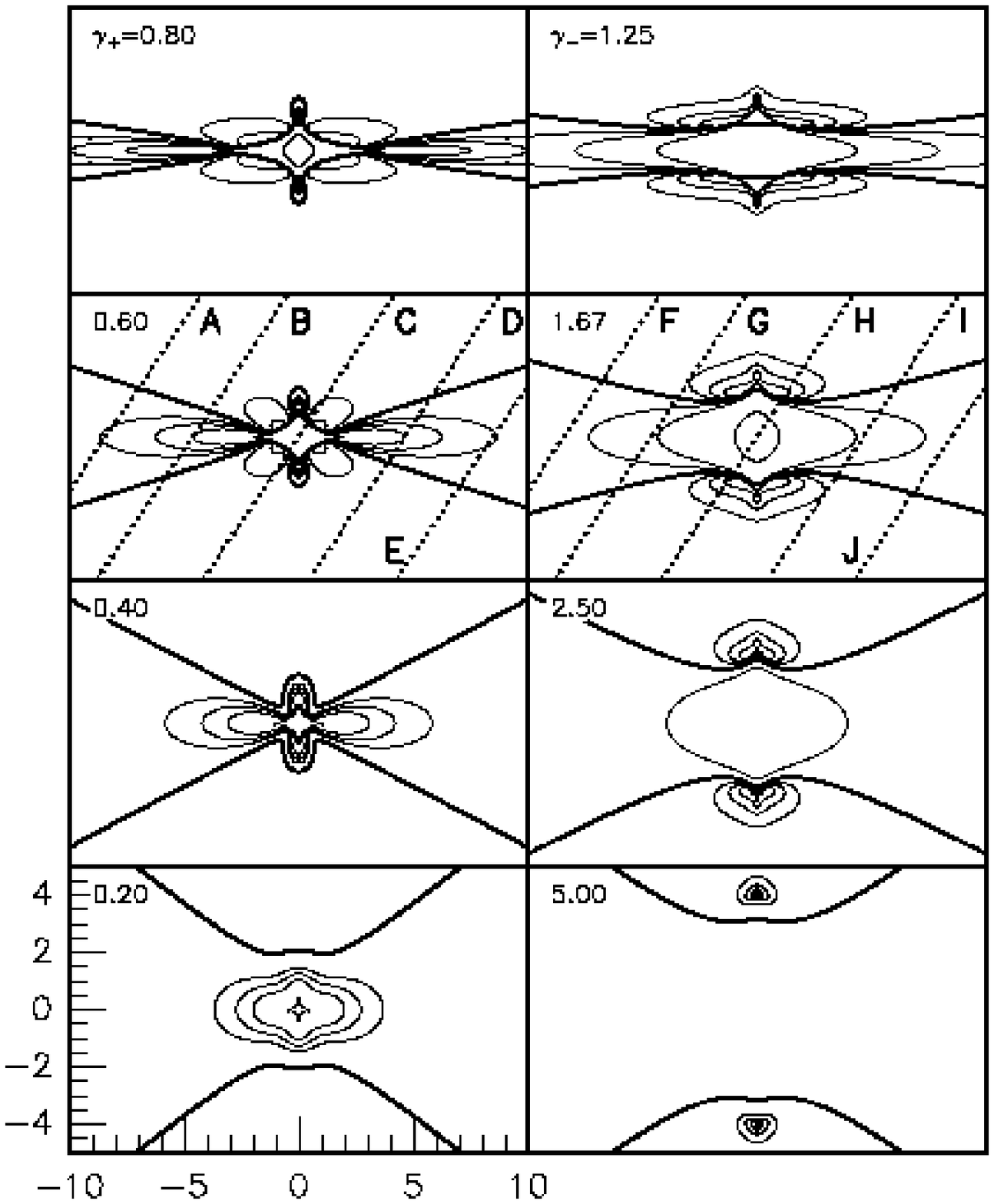}{6.4}{6.5}{7.0}
{
Chang-Refsdal magnification contours of a point source as
a function of source position in units of the planet Einstein ring, $\theta_p=q\theta_e$,
for various pairs of shears
$(\gamma_+$ and $\gamma_-=\gamma_+^{-1})$ corresponding to planetary 
perturbations 
of the major and minor images, respectively. 
Magnification contours are calculated including the contribution
of the unperturbed image. Contour pairs are for
$\gamma_+=0.8,$ 0.6, 0.4, and 0.2 in panels (a,b), (c,d), (e,f), and (g,h),
and correspond to source positions at perturbation of 
$x_d = 0.22,$ 0.52, 0.95, and 1.79.  Super-bold contour is no deviation.
Bold contours are $\delta = 5\%$, 10\%, 20\%, and $\infty$.  Non-bold contours
are $-5\%$,  $-10\%$,  and $-20\%$.  Diagonal lines in panels (c) and (d)
represent possible trajectories assuming that the overall light curve shows
$x_d=0.52$ (i.e.\ $\gamma_+=0.6$) and $\beta=0.4$ (i.e.\ $\phi=\sin^{-1}{\beta/
x_d} \sim 50^\circ$).
If the maximum deviation were observed to be $\delta_d=20\%$ (and the 
point-source approximation were known to be valid), then the trajectory
must be either B or D. 
}
\endinsert

	The Chang-Refsdal approximation permits an immense conceptual 
simplification of the problem.  For a point source, all possible 
light curves of an event with a given $x_d$ can be represented on a 
pair of diagrams, one for $\gamma_+$ and one for $\gamma_-$.  All possible
planetary perturbations can therefore be represented by a single-parameter 
family of such diagrams.  
See Figure \three.  For a given event, one knows $\beta$
and $x_d$ from the overall light curve.  One can therefore compute 
$\gamma_\pm$ using equation \gammadef\ and thereby pick out which two diagrams
are relevant.  One also knows the angle $\phi$ at which the source cuts through
the diagram,
$$\sin \phi = {\beta\over x_d}.\eqn\sinphi$$
If, for example, $x_d=0.516$ and $\beta=0.4$, then all possible light curves
are represented by the parallel lines indicated in Figures \three-c,d.  If
the light curve is well sampled, it is easy to distinguish between $\gamma_+$
and $\gamma_-$.  Suppose that $\gamma_+$ (Fig.\ \three-b) is correct, and
say that the maximum fractional deviation is $\delta_d=20\%$.
Then one can immediately identify the correct curve as being either B or D.
The observed duration of the perturbation relative to that of
the whole event then
sets the scale of diagram relative to $\theta_e$ and thus determines the
mass ratio.

\FIG\four{
Chang-Refsdal magnification contours of a non-point source as
a function of source position in units of $\theta_p$, for the pair 
of shears 
$(\gamma_+,\gamma_-)=(0.6,1.67)$ corresponding to planetary perturbations 
at source position $x_d=0.52$.  The ratios of planet Einstein radius to
source radius are $\rho=0.5,$ 1.0, 1.5, and 2.5.  Compare with point-source
case shown in Figs.\ \three-c,d.  Contours levels are the same as for Fig.\ 
\three.
}

\topinsert
\mongofigure{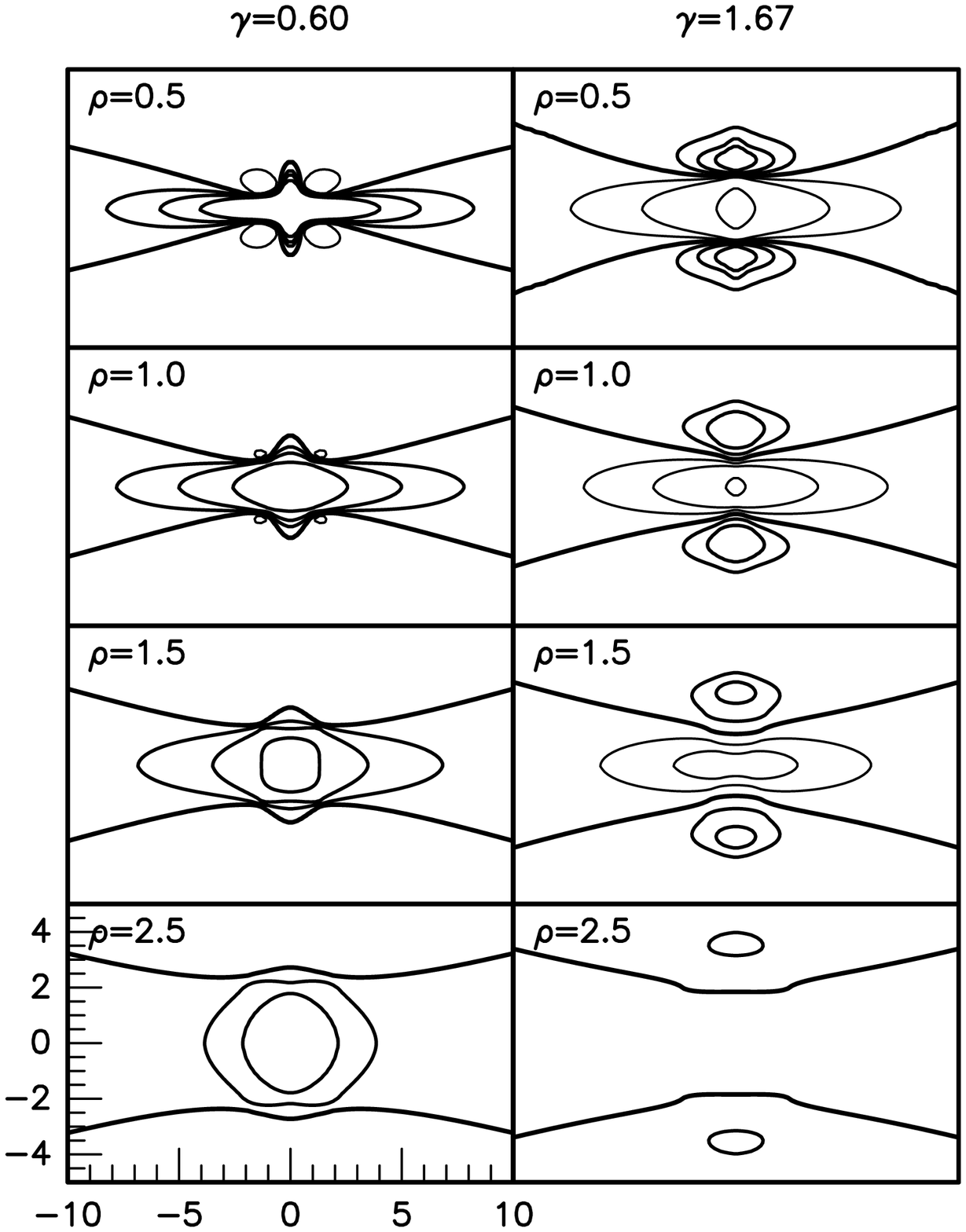}{6.4}{7.0}{7.0}
{
Chang-Refsdal magnification contours of a non-point source as
a function of source position in units of $\theta_p$, for the pair 
of shears 
$(\gamma_+,\gamma_-)=(0.6,1.67)$ corresponding to planetary perturbations 
at source position $x_d=0.52$.  The ratios of planet Einstein radius to
source radius are $\rho=0.5,$ 1.0, 1.5, and 2.5.  Compare with point-source
case shown in Figs.\ \three-c,d.  Contours levels are the same as for Fig.\ 
\three.
}
\endinsert

	Of course, one does not know {\it a priori} that finite source
effects can be ignored.  However, for any $x_d$, all possible events can still
be represented by a single-parameter family of diagrams.  The relevant 
parameter is, $\rho$,
the ratio of the angular radius of the source to the Einstein radius of the
planet.  Hence, it is quite easy to study all possible degeneracies.
See Figure \four.

	The drawback of using the Chang-Refsdal approximation is that it is
not exact.  Moreover, for any given lensing event, it is straight forward to
construct models that are exact.  As we argue below, however, the lack of
exactness has no significant impact in the analysis of degeneracies.  On the
other hand, using the exact solution increases the dimensionality of parameter
space and thereby the conceptual complexity of the problem, without any
compensating benefits.  We therefore strongly advocate using the Chang-Refsdal
framework.  We present a more detailed analysis in the Appendix.

\chapter{Degeneracy Between Major and Minor Images}

	Figure \three\ indicates that it should generally be quite easy to
distinguish between perturbations of the major and minor images provided that
there is good temporal coverage: perturbations of the major image have one
major positive excursion, while perturbations of the minor image have two
positive excursions separated by a large negative excursion.  However, if the 
observations are made from only one site, then good temporal coverage is
far from automatic.  The time scale for these excursions is the minimum
of the crossing time of the star, $\theta_*/\mu\sim 10\,$ hours for a giant,
and the crossing time of the planet Einstein ring $\theta_p/\mu\sim 
10\,(m/50 M_\oplus)^{1/2}\,$hours.  Thus it would be quite possible to observe
a positive excursion (or a significant fraction of it) one night and then
miss any subsequent excursions due to one or two nights of bad weather.
However, if there were three observing sites on different continents, such
large data gaps would be rare.

\FIG\five{
Major/minor image degeneracy for $\phi=0$, with $\gamma_+=0.61, 0.38, 0.25$,
in panels a,c,and e. The solid curve corresponds to $\gamma_+$, the
dashed curve to $\gamma_-=\gamma_+^{-1}$. Also shown in panels b,d, and f are
the associated fractional color change $\Delta(V-H)$.
}

\topinsert
\mongofigure{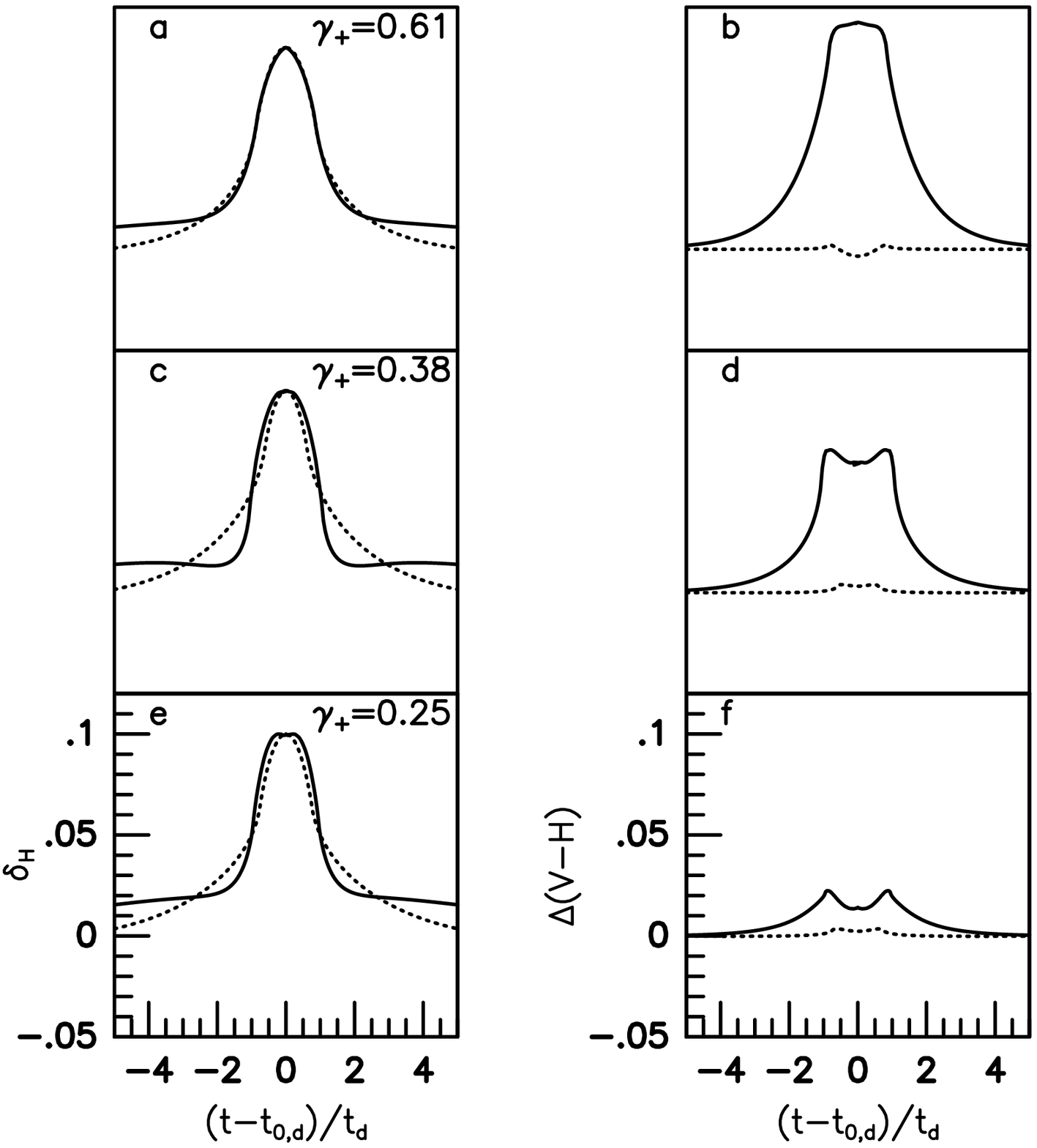}{6.4}{6.5}{6.5}
{
Major/minor image degeneracy for $\phi=0$, with $\gamma_+=0.61, 0.38, 0.25$,
in panels a,c,and e. The solid curve corresponds to $\gamma_+$, the
dashed curve to $\gamma_-=\gamma_+^{-1}$. Also shown in panels b,d, and f are
the associated fractional color change $\Delta(V-H)$.
}
\endinsert

	There is nevertheless another possible source of  degeneracy between
the major and minor images.  If $\phi$ is sufficiently small, then a source
coming close to one of the caustics of the perturbation of the minor image
could cross the star-planet axis at a point far enough from the planet that
the negative excursion along this axis would be very small.  In this case,
the light curve might be mistaken for one due to a perturbation
of the major image.  In Figure \five, we present examples of this
degeneracy for three different values of $\gamma_+$ for $\phi = 0$,
along with the corresponding color shifts $\Delta(V-H)$ (see \S\ 6.3).
The parameter $\rho$ was chosen in each case such that the 
curves for $\gamma_+$ and $\gamma_-$ would be most similar.     
It is clear that
the degenerate curves could be distinguished only if precise measurements
could be made at the wings of the perturbation. 
Considering now the curves for 
$\Delta(V-H)$, there are relatively large ($2-10\%$) fractional
color changes associated with the $\gamma_+$ curves throughout the event, 
while the fractional color changes
associated with the $\gamma_-$ curves are always negligible. 
This large difference in the magnitude of the fractional color change
between the $\gamma_+$ and $\gamma_-$ curves arises from the fact that
the gradient of the magnification across the face of the star is 
much larger in the $\gamma_+$ case, and thus the color effects are 
therefore more pronounced (see \S\ 6.3). 
Thus by measuring $\Delta(V-H)$, one can 
distinguish between the two degenerate cases
even during the peak of the perturbation. 
The larger the value of $\phi$, the larger the negative excursion in the
minor-image perturbation, and thus curves
 with $\phi > 0$ will be less degenerate than the examples shown
in Figure \five. 

\chapter{Degeneracy of Planet Position Relative to Unperturbed Image}

	In general, this degeneracy introduces an uncertainty in $y_p$
which is $\Delta y_p\sim 2\alpha\theta_p/\theta_e$ where $\alpha\theta_p$ 
is the separation between the planet and the unperturbed image at the mid-point
of the perturbation.  From Figure \three\ one sees that if perturbations
$\delta_d\sim 5\%$ are detectable, the typical planetary event will have
$\alpha\sim 5$.  Hence, if the degeneracy remains unbroken, the fractional
uncertainty is $\Delta y_p/y_p\sim (100 m/M)^{1/2}$.  For Jupiter-mass
planets in orbit around M dwarfs, this error is of order unity, while 
for Neptune-mass planets it is $\sim 10\%$.  On the other hand, the degeneracy
in $q$ is small (see Table 1 and also Appendix).  

	In this section we consider only point sources.  If the planet has
a low mass, so that finite source effects are important, then (as mentioned
above) the difference between the two degenerate solutions is small and
distinguishing between them is relatively less important.  In addition,
finite-source effects are more properly addressed in the context of the
mass/finite-source degeneracy discussed in \S\ 6.

\section{Perturbations of the Major Image}

\FIG\six{
Asymmetry factor $P$ (see eqs.\ \pdef\ and \pphidef) for nine values
of $\gamma$, 
as a function of maximum deviation, $\delta_d$.  The actual asymmetry
of a given light curve, $P_\phi(\gamma,\delta_d)$, is given by 
$P_\phi(\gamma,\delta_d)\sim  P(\gamma,\delta_d)\cot\phi$.  Using this figure
and formula, one can therefore determine whether the degeneracy can be
broken for any given sensitivity threshold.
}

	It is clear from Figure \three\ that it is impossible to break the
degeneracy if $\phi=90^\circ$, i.e., if the planetary perturbation takes
place at the peak of the light curve ($x_d=\beta$).  If the source crosses 
the perturbation structure in the region of the caustic $(\alpha\lsim 1)$
then degeneracy is relatively unimportant and, in any event, is easily
broken (provided $\phi<90^\circ$) due to the richness of the structure
in this region.  We therefore
focus on the case $\alpha>1$.  From Figure \three, one
sees that there is an asymmetry in the light curve which has opposite senses
depending on whether the source crosses to the left or the right of the 
planet.  If it passes to the right, then the deviation is more pronounced
at the beginning of the perturbation than at the end, and if it passes to
the left, the deviation is more pronounced at the end.  
We define the
asymmetry factor $P_\phi$ as the maximum over all times $t$ of the fractional 
difference,
$$ P_\phi(\gamma,\delta_d) \equiv {\max\{|\delta(t_{0,d} + t)-
\delta(t_{0,d} -t)|\}}\eqn\pdef$$
where $t_{0,d}$ is the mid-point of the perturbation and
$\delta(t)$ is the fractional 
deviation as a function of time.  To lowest order,
one may approximate 
$$P_\phi(\gamma,\delta_d)= P(\gamma,\delta_d)\cot\phi .\eqn\pphidef$$
From Figure \three, one can see that as $\gamma$ increases, the
positive contours of $\delta$ become more stretched along the
planet-star axis, and thus low-peak perturbations occur farther
from the areas of negative excursion.  One would therefore expect smaller
values of $P$ for larger
values of $\gamma$.   
Figure \six\ shows $P$ as a function of $\delta_d$ for several values of
$\gamma$.    
As expected, $P$
generally decreases with increasing $\gamma$.  From
Figure \six, one can determine, for a given sensitivity,
whether the degeneracy can be broken for any perturbation.  
For example, consider trajectories
such that $\phi \gsim 75^{\circ}$. If one
were sensitive to asymmetries of $P_{\phi} \sim 1\%$, i.e., 
$P \sim .04$, then the degeneracy could
be broken only for $\gamma \lsim 0.3$ , and then only if
$\delta_d > 0.1$.  On the other hand, if $\phi \sim 30^{\circ} (P\sim 0.004)$,
then the degeneracy could be broken for essentially all values of $\gamma$.

\topinsert
\mongofigure{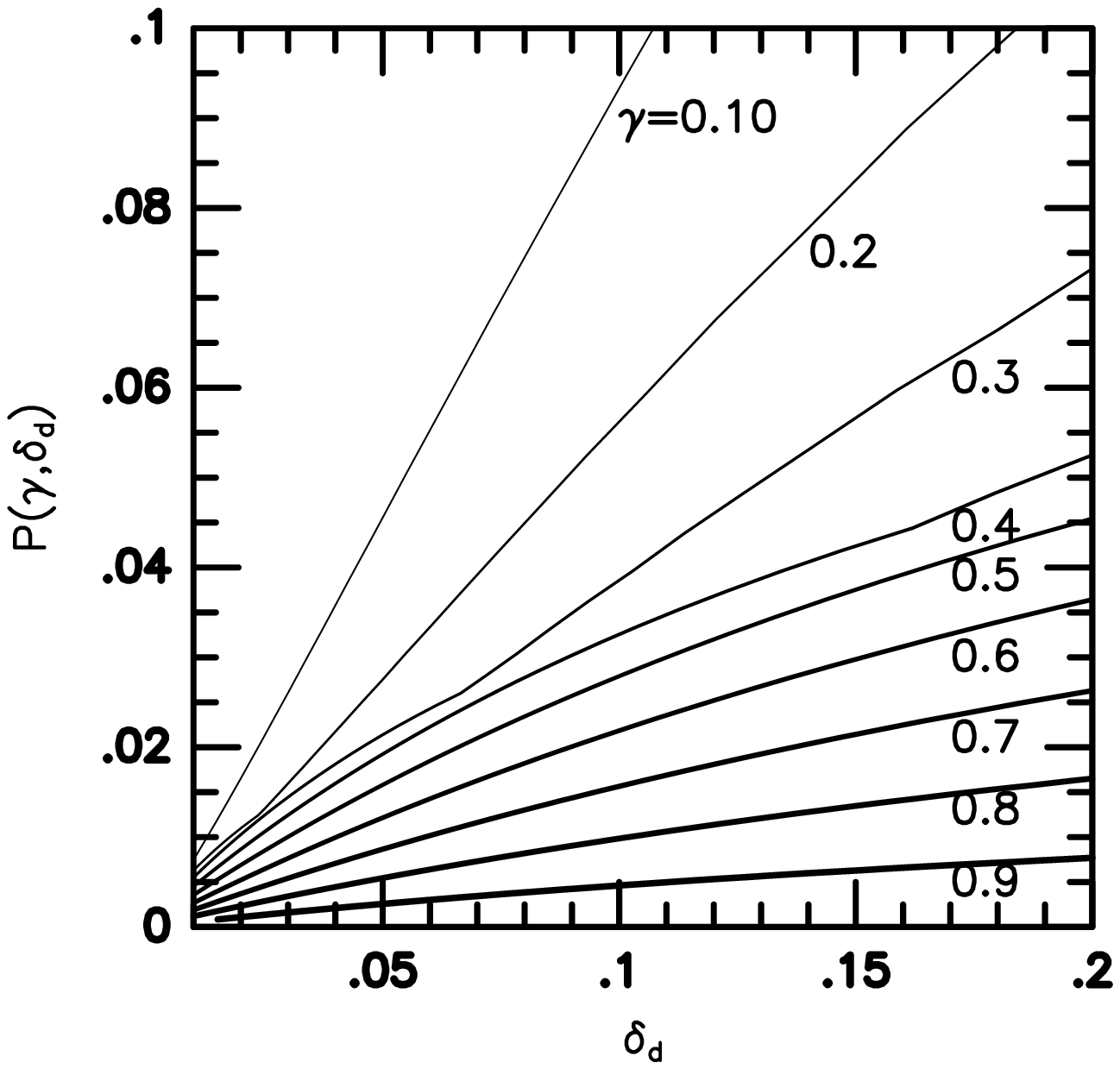}{6.4}{5.0}{6.5}
{
Asymmetry factor $P$ (see eqs.\ \pdef\ and \pphidef) for nine values
of $\gamma$, 
as a function of maximum deviation, $\delta_d$.  The actual asymmetry
of a given light curve, $P_\phi(\gamma,\delta_d)$, is given by 
$P_\phi(\gamma,\delta_d)\sim  P(\gamma,\delta_d)\cot\phi$.  Using this figure
and formula, one can therefore determine whether the degeneracy can be
broken for any given sensitivity threshold.
}
\endinsert

\section{Perturbations of the Minor Image}

	As we discussed in \S\ 4, to distinguish minor-image from 
major-image perturbations, it is necessary to observe the negative excursion
(centered on the $x$-axes of the right-hand side of Fig.\ \three.   If these
are observed, then one can easily distinguish the case where the source
transits the right side from the case where it transits the left side of the
$x$-axis (provided $\phi<90^\circ$).  Hence, there is no degeneracy of 
minor-image perturbations unless the more severe major/minor-image degeneracy
remains unbroken.

\chapter{Continuous Mass Degeneracy of Major Image Perturbations}

	By far the potentially most crippling form of parameter degeneracy
is the one that is illustrated in Figure \two\ and is tabulated in Table 2.
The basic character of this degeneracy can be understood analytically using 
the following 
theorem (Gould \& Gaucherel 1996): if the unperturbed major image
crosses the position of the planet and the source is larger than
the major-image caustic structure, then
$$\delta_d \simeq {2\over \rho^2 A(\gamma)},\qquad A(\gamma) = 
{1+\gamma^2\over 1-\gamma^2}.\eqn\massdegone$$
The FWHM of such an event is $t_d\sim 2(\csc\phi)\rho q^{1/2}t_e$.  On the
other hand, the FWHM of a low-peak perturbation of a point-source is
$t_d\sim 2(\csc\phi) q^{1/2}t_e$.  Suppose that an event has observables
$t_e$, $\beta$, $x_d$, $t_d$, and $\delta_d$.  One can form the combination
of observables $Q\equiv [(\beta/x_d)(t_d/t_e)/2]^2$, and can obtain one
possible solution that reproduces the maximum deviation and FWHM:
$$ q\sim Q,\qquad \rho\lsim 1.\eqn\solone$$
However, the solution
$$ q\sim  {Q\over \rho_\max^2}, \qquad \rho \sim \rho_\max,  \qquad \rho_\max
\equiv \biggl({2\over\delta_d}\,
{1-\gamma^2\over 1+\gamma^2}\biggr)^{1/2},\eqn\soltwo$$
would also reproduce the height and width of the curve.  Note that the
ratio of masses for the two solutions is $\rho_\max^2$.  For 
$\delta_d\sim 5\%$, this ratio is typically $\gsim 20$ and can be as high
as 40.  Thus, unless this degeneracy is broken, any low-peak perturbation
of a point source by a Jupiter-mass planet can masquerade as a Neptune-mass
event, and vice versa.  All intervening masses are permitted as well.
Clearly, unless this degeneracy is broken, low-peak perturbations will contain
very little information about mass, and unambiguous detection of low-mass
planets will be impossible.  There are three possible paths to breaking this 
degeneracy.
\FIG\seven{
Fractional deviations $\delta$ for $H$-band light curves.  
Similar to Fig.\ \two, except now shown for shears 
$\gamma=0.2,$ 0.4, 0.6, and 0.8, and for trajectory of source motion
$\phi=30^\circ$, $45^\circ$, $60^\circ$, and $90^\circ$.  (Corresponding curves
for $\phi\rightarrow -\phi$ can be found by reversing the $x$-axes.)\ \ 
In each curve, the maximum
deviation is $\delta_d=10\%$ and the FWHM is $t_d=0.06\,t_e$.  
}
\FIG\eight{
The values of $(q/q_0)^{-1}$ (bold lines) and $(\mu/\mu_0)^{-1}$ 
as functions of
$\rho$ for each set of degenerate curves in Fig.\ \seven.  The
fiducial values $q_0$ and $\mu_0$ are associated with the curve
with $\rho=0.1$.
}
\FIG\nine{
Fractional color change $\Delta(V-H)$ for light curves shown in Fig.\ \seven.
}

\topinsert
\mongofigure{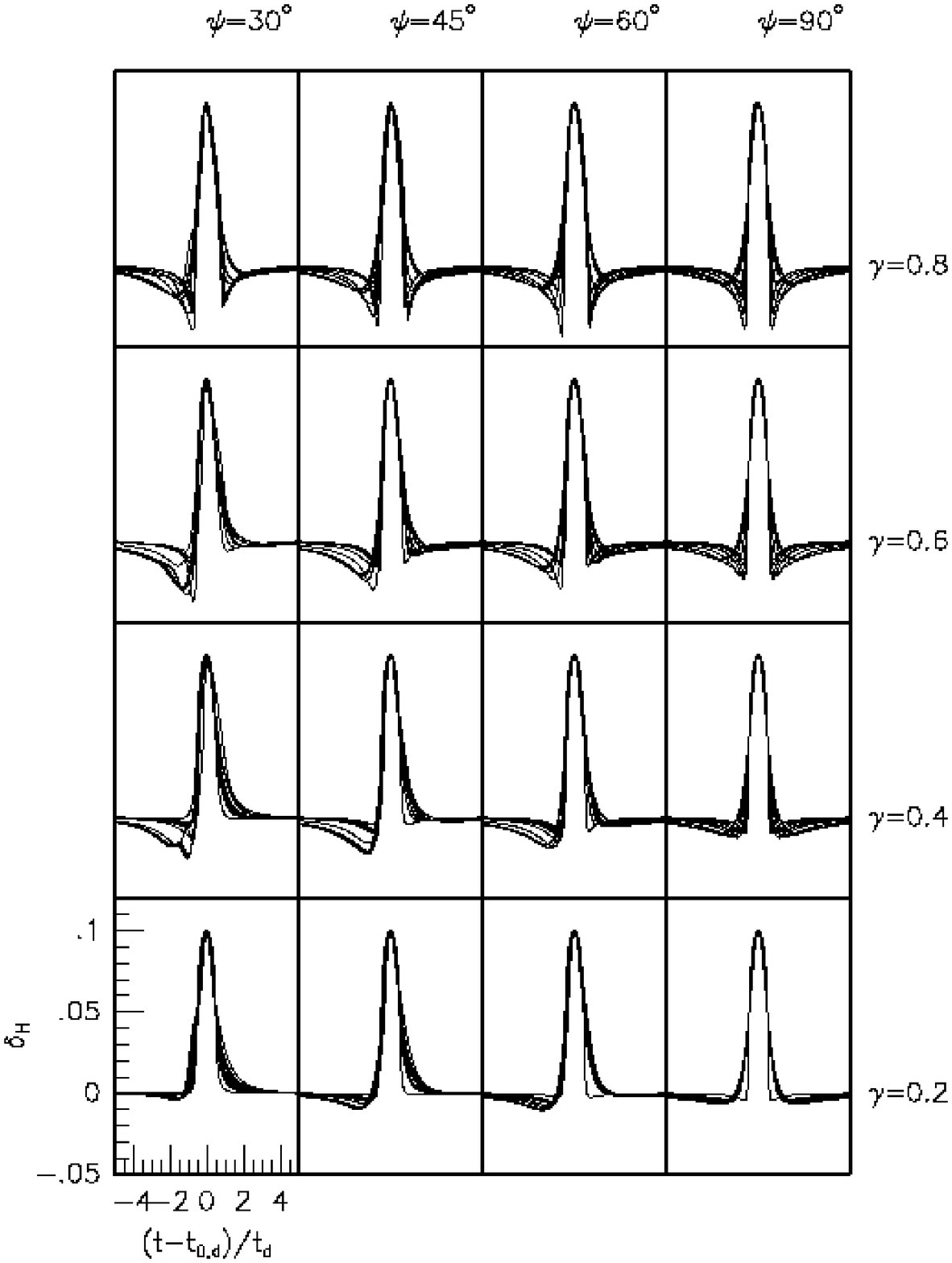}{7.5}{8.5}{7.5}
{
Fractional deviations $\delta$ for $H$-band light curves.  
Similar to Fig.\ \two, except now shown for shears 
$\gamma=0.2,$ 0.4, 0.6, and 0.8, and for trajectory of source motion
$\phi=30^\circ$, $45^\circ$, $60^\circ$, and $90^\circ$ (left to right).  
Corresponding curves
for $\phi\rightarrow -\phi$ can be found by reversing the $x$-axes.\ \ 
In each curve, the maximum
deviation is $\delta_d=10\%$ and the FWHM is $t_d=0.06\,t_e$.
}
\endinsert

\topinsert
\mongofigure{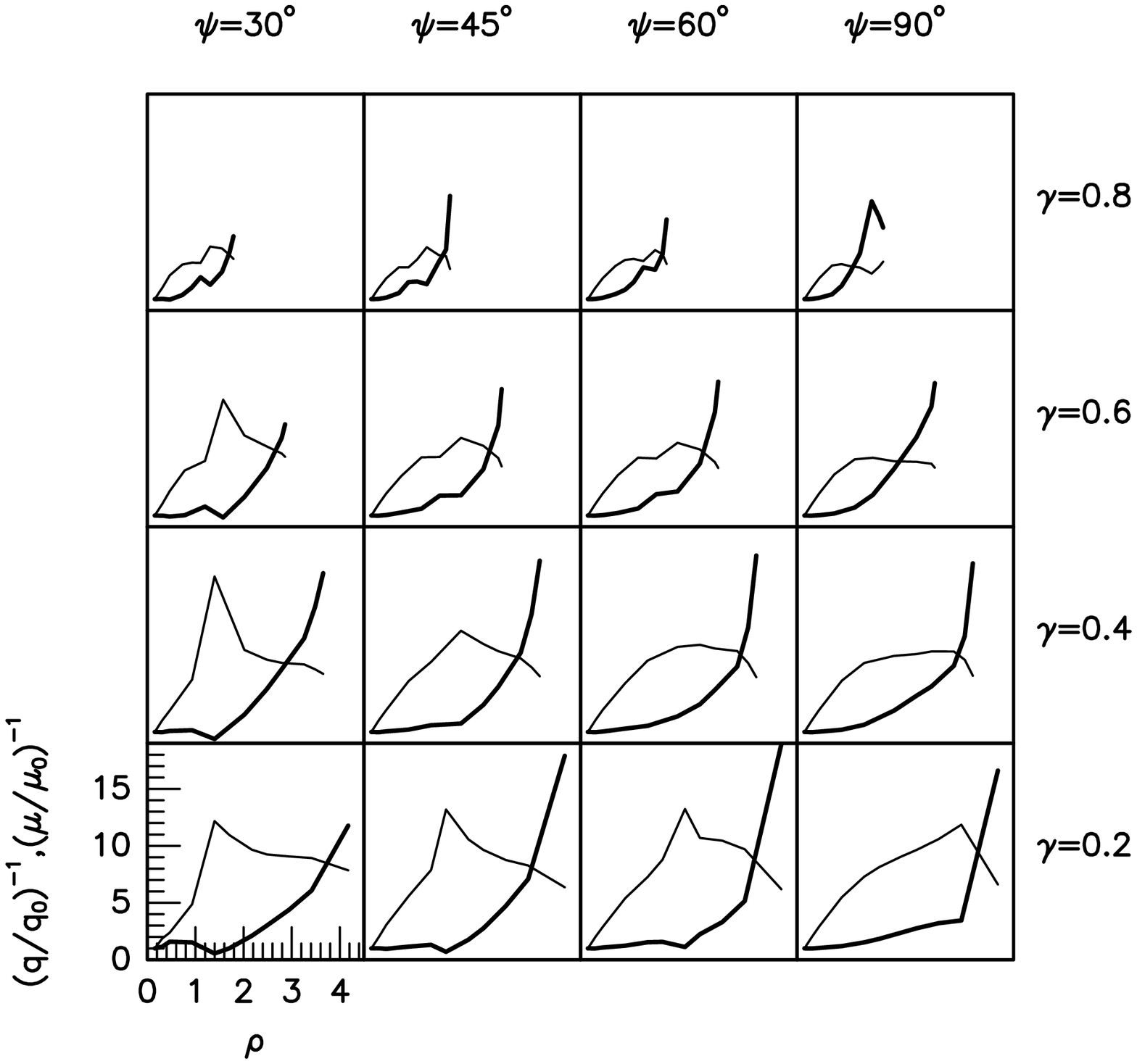}{7.5}{7.5}{7.5}
{
The values of $(q/q_0)^{-1}$ (bold lines) and $(\mu/\mu_0)^{-1}$ 
as functions of
$\rho$ for each set of degenerate curves in Fig.\ \seven.  The
fiducial values $q_0$ and $\mu_0$ are associated with the curve
with $\rho=0.1$.
}
\endinsert
\topinsert
\mongofigure{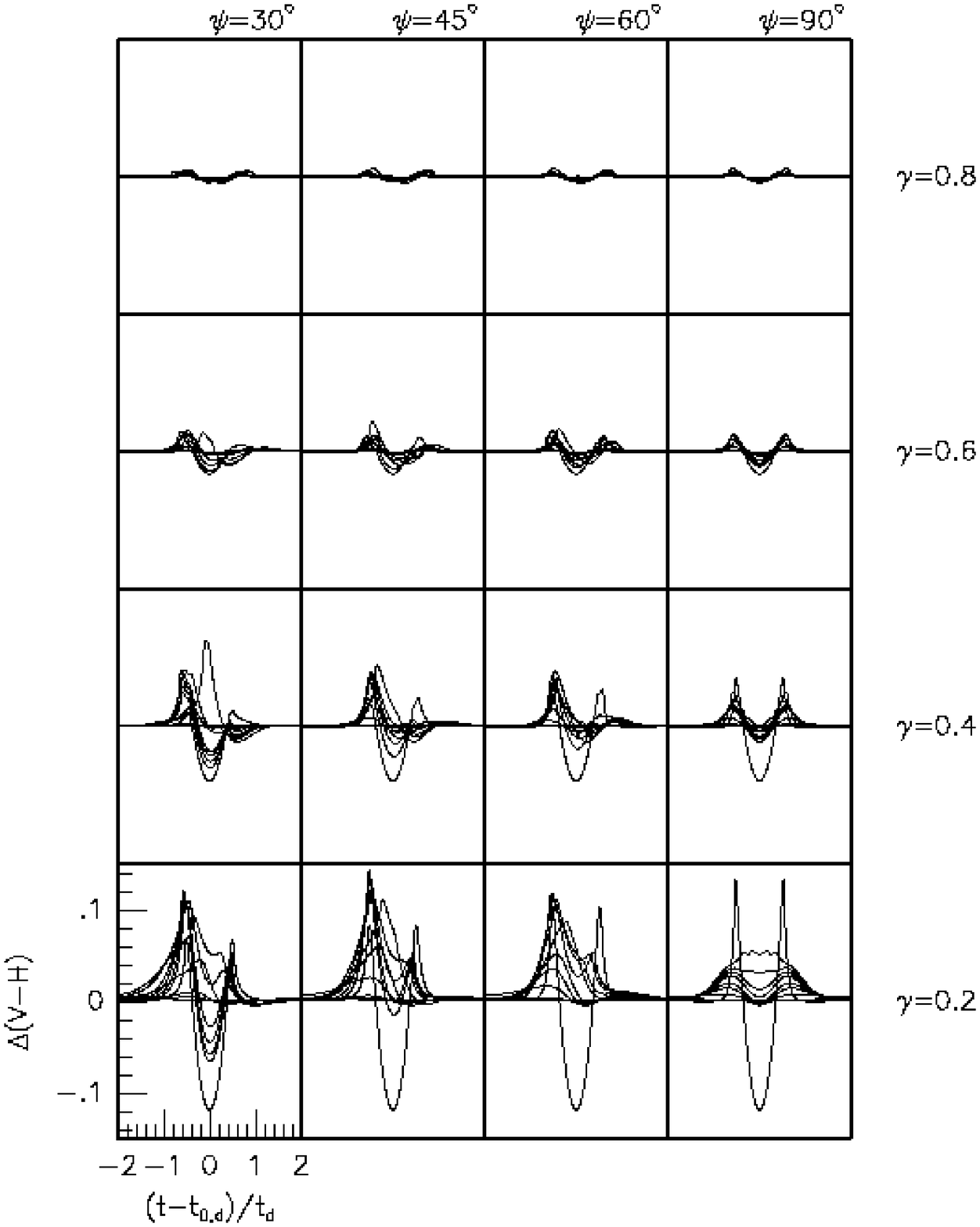}{9.0}{8.7}{7.5}
{
Fractional color change $\Delta(V-H)$ for light curves shown in Fig.\ \seven.
}
\endinsert

\section{Proper Motion Measurement}

	If the proper motion $\mu$ of the lens relative to the source were 
measured, then one could partially break the degeneracy between equations 
\solone\ and \soltwo.  The angular size of the source $\theta_*$ would be
known from its dereddened color and magnitude and Stefan's law.  The time
for the source to cross the perturbation $x$-axis, $t_c\sim 2\theta_*\csc\phi/\mu$,
would then also be known.  If one found $t_c<t_d$, this would imply that
$t_d$ was dominated by the size of the planet Einstein ring, not the source.
Hence, the solution \solone\ would be indicated.  On the other hand, if
$t_c\sim t_d$, one would
 know only that the solution \solone\ was not correct, and
that the mass ratio lay somewhere in the interval $\rho_\max^{-2}Q\leq q<Q$.
See Table 2.

	Proper motions can be measured if the lensing star transits or nearly 
transits the source or by imaging the split image of the source using
infrared interferometry.  See Gould (1996) for a review.  Another approach
would be simply to wait a few decades and measure the angular separation of
the lens and source.  Since typical proper motions are 5 mas yr${}^{-1}$, the
separation should be $\sim 0.1''$ after a few decades.  Unfortunately, most
lenses are probably fainter than $M_I=10$, while typical giant sources are
$M_I\sim 0$, and even turnoff stars are $M_I\sim 3$.  Thus it would be 
difficult to image the lens until it was quite well separated from the source 
at which point it might be hard to distinguish it from random field stars.  
We explore this possibility further in \S\ 8.
 
\section{Detailed Light Curves}

	Although the parameter combinations \solone\ and \soltwo\ reproduce
the gross features of the perturbation (peak and FWHM) equally well, the
detailed structures of the light curves are different.  Figure \two\ 
illustrates the principal difference for elongated perturbation
structures, in this case $\gamma=0.6$.  When $\rho\sim \rho_\max$, the wings 
show a dip because the source passes over the caustic which is surrounded
by regions of negative perturbation (see Figs.\ \three-c and \four).  On
the other hand, when $\rho\lsim 1$ the approach to the peak is smooth
because the source is passing over the smooth outer portion of the ridge
seen in Figure \three-c.  From Figure \two\ it is clear that
if these wing structures could be resolved at the $\sim 1\%$ level, then
the degeneracy in mass could be reduced from the factor $\sim 15$ seen
in Table 2, to a factor $\sim 1.5$.

	Figure \seven\ is an array of 16 diagrams each similar to Figure \two,
but with different values of $\gamma$ (0.2, 0.4, 0.6, and 0.8) and different
angles of source motion $\phi$ ($30^\circ$, $45^\circ$, $60^\circ$,
and $90^\circ$).
It is clear  
that it is easier to break the degeneracy as $\gamma$ 
increases and as $\phi$ decreases.  For $\gamma \gsim 0.4$
the uncertainty in $q$ could be significantly reduced if
the wing structures could be resolved at the $\sim 1\%$ level. 
For $\gamma \lsim 0.2$, however, 
distinguishing between the degenerate curves would
require an accuracy $\ll 1\%$.  
Figure \eight\ shows the values of $q/q_0$ and $\mu/\mu_0$ as a function
of $\rho$ for each of the combinations of $\phi$ and $\gamma$ in Figure
\seven.  Note that larger values of $\rho_\max$ are allowed for
smaller values of $\gamma$, and thus the range of acceptable values of
$q$ is largest for small $\gamma$.  This is especially disturbing in
light of the fact that the degenerate curves are most similar for
small $\gamma$.  

\section{Optical/Infrared Colors}

	A major shortcoming of the detailed-light-curve method for 
breaking the degeneracy is that it depends critically on obtaining accurate
observations during two brief intervals covering the wings of the light curve.
As a practical matter, it may be difficult to obtain such coverage for
a variety of reasons.  Once the event is noticed, observatories that are 
dedicated to the planet search can engage in frequent monitoring and thereby
obtain very accurate light curves.  However, it is quite possible, indeed
likely, that the planetary perturbation will not be recognized in time for
intensive monitoring of the first wing.  Sometimes observation of the first
wing is crucial to breaking the degeneracy.  Moreover, the second wing will 
likely be observable from at most one observatory which could be affected 
by bad weather.

	Optical/infrared color measurements by contrast yield 
degeneracy-breaking information throughout the event.  The reason is that
by the principle of equivalence, lensing of point source is achromatic.  If
lensing introduces color changes, the lens must be resolving the source
(Witt 1995; Loeb \& Sasselov 1995).  The best opportunity to observe this
effect is by looking for optical/infrared color differences 
(Gould \& Welch 1996) because giant
stars are more limb-darkened in the optical than in the infrared 
(Manduca, Bell \& Gustafsson 1977).
Thus, if the planet Einstein ring is larger than the source (and the low peak
is due to the source passing over regions of small perturbation), the color
changes will be very small.  On the other hand, if $\theta_*>\theta_p$
(and the low peak occurs when the large source passes over the caustic), the
caustic structure will resolve the differential limb-darkening of the star
and the color changes will be more pronounced.
Figure \nine\ shows the $V-H$ colors for the same parameters as are used for
the $H$ band curves as in Figure \seven.  The magnitude of the 
fractional color change is largest for smallest $\gamma$.  This is 
fortunate, since, as discussed in \S\ 6.2, the degeneracy is most severe
for small $\gamma$, both in terms of the similarity in the light curves,
and in the range of allowed values of $q$.  It is therefore essential
to have optical/infrared color measurements to ensure that the
continuous degeneracy can be broken for all possible values of $\gamma$.

\chapter{Continuous Degeneracy of Minor Image Perturbations}

\FIG\ten{a) Ten light curves with $\gamma_-=1.67$, $\phi=90^\circ$, 
all with maximum 
deviation $\delta_d=-10\%$ and FWHM
$t_d = 0.06\,t_e$.  The ratios of source radius to planet Einstein ring range
from $\rho=0.1$ to $\rho=1.83$, the largest source radius consistent
with this maximum deviation.  The corresponding relative
values of $q=m/M$, and the
relative proper motion, $\mu/\mu_0$, are given in Table 4.
Source radii of $\rho = 1.6, 1.7, 1.83,$ and $1.80$ correspond
to the bold, dashed, bold dotted, and bold dashed curves.   
b) The fractional color change $\Delta(V-H)$ for the ten curves in panel (a).
}

There is also a continuous degeneracy for minor-image perturbations,  
but the degeneracy is considerably
less severe than for major-image perturbations because
the caustic structure is qualitatively different.
As with major image perturbations, the basic character of the minor
image degeneracy can be understood analytically.  Consider the following
theorem (Gould \& Gaucherel 1996): if the unperturbed minor image
crosses the position of the planet and the source encloses
both minor-image caustics,  then
$$
\delta_d \simeq -{{f(\rho,\gamma)}\over \rho^4A(\gamma)}
\rightarrow -{2\over{\rho^4 A(\gamma)}},\qquad A(\gamma) = 
{\gamma^2+1\over \gamma^2-1},\eqn\massdegtwo
$$
where $f(\rho,\gamma)=[(1/2 + \rho^{-2})^2 -\gamma^2\rho^{-4}]^{-1/2}$,
and the limit applies for $\rho \gg \gamma^{1/2}$.
That is, in contrast to the major-image perturbation [cf.\ eq.\ \massdegone],
the minor-image perturbation goes to zero rapidly for large sources.
For minor image perturbations, the caustics are located at
(Schneider et al.\ 1992)
$$d_{\rm{caus}} \sim 2(\gamma-1)^{1/2}. \eqn\caus$$
Thus the source must have $\rho \gsim 2(\gamma-1)^{1/2}$
to enclose both caustics.  If the source is significantly
larger than this, the perturbation will be negligibly small.
Hence we can restrict attention to sources $\rho < d_{\rm{caus}}$.

For minor-image perturbations, $t_d$ is the FWHM of 
the negative deviation.  
For a low-peak point source perturbation ($\alpha \gsim 2$), 
this duration scales as the distance between the contours of $\delta = 0$ 
at $\alpha$. (See Fig.\ \three).
For finite sources of increasing $\rho$, the trajectories must move
closer to the center to maintain the observed value of $\delta_d$.
(See Fig.\ \four).  However, until $\rho$ becomes so large as to cover
the caustics, the positions of the $\delta=0$ contours basically 
do not change. (Compare Fig.\ \three d with Fig.\ \four).  Since
these contours are approximately horizontal, $t_d$ is not greatly
influenced by changes in $\rho$.  Finally, since the largest permitted
source has $\rho \sim d_{\rm{caus}}$, which is also approximately
equal to the separation of the $\delta=0$ contours, $t_d$ remains roughly
the same even for this extreme case (see Fig.\ 3).
The small degeneracy that does exist arises from the difference 
between this extreme case on the one hand and the smaller sources
and point sources on the other.  Examining Figure 3, we can expect that the 
degeneracy will be somewhat larger for larger values of $\gamma$, since
the contours of $\delta=0$ become less horizontal as $\gamma$ increases.
In Table 3 we give the degeneracy in the inferred values of $q$ for 
$\delta_d = -10\%$ and $-5\%$, and for several 
values of $\gamma$.  Note that  
the largest degeneracy in $q$ is only a factor of $\sim 4$.

$$\vbox{\halign{#\hfil\quad&\hfil#\quad&\hfil#
\quad&\hfil#\hfil\quad&\hfil#\hfil\quad&\hfil#\hfil\cr
\multispan{3}{\hfil TABLE 3 \hfil}\cr
\noalign{\medskip}
\multispan{3}{\hfil 
Continuous Minor Image Degeneracy\hfil}\cr
\noalign{\smallskip}
\noalign{\hrule}
\noalign{\smallskip}
\noalign{\hrule}
\noalign{\smallskip}
\hfil \hfil&\hfil\hfil&\hfil mass ratio \hfil\cr
\hfil \hfil&\hfil\hfil&\hfil degeneracy\hfil\cr
\hfil $\delta_d$\hfil&\hfil $\gamma$\hfil&\hfil $q_{\rm{max}}/q_{\rm{min}}$ \hfil\cr
\noalign{\smallskip}
\noalign{\hrule}
\noalign{\smallskip}
\hfil-10\%\hfil&\hfil1.25\hfil&\hfil1.45\hfil\cr
\hfil     \hfil&\hfil1.43\hfil&\hfil1.09\hfil\cr
\hfil     \hfil&\hfil1.67\hfil&\hfil1.68\hfil\cr
\hfil     \hfil&\hfil2.00\hfil&\hfil3.82\hfil\cr
\noalign{\smallskip}
\noalign{\hrule}
\noalign{\smallskip}
\hfil -5\%\hfil&\hfil1.25\hfil&\hfil1.66\hfil\cr
\hfil     \hfil&\hfil1.43\hfil&\hfil1.16\hfil\cr
\hfil     \hfil&\hfil1.67\hfil&\hfil1.22\hfil\cr
\hfil     \hfil&\hfil2.00\hfil&\hfil1.67\hfil\cr
\hfil     \hfil&\hfil2.50\hfil&\hfil2.34\hfil\cr
\noalign{\smallskip}
\noalign{\hrule}
}}
$$

\FIG\ell{Ten light curves with $\gamma_-=0.6^{-1}$, $\phi=60^\circ$, 
all with maximum deviation $\delta_d=-10\%$ and FWHM
$t_d = 0.06\, t_e$.  All other parameters are the same
as Fig. \ten, and are given in Table 4.  Curves are
as in Fig. \ten.
}
 \topinsert
\mongofigure{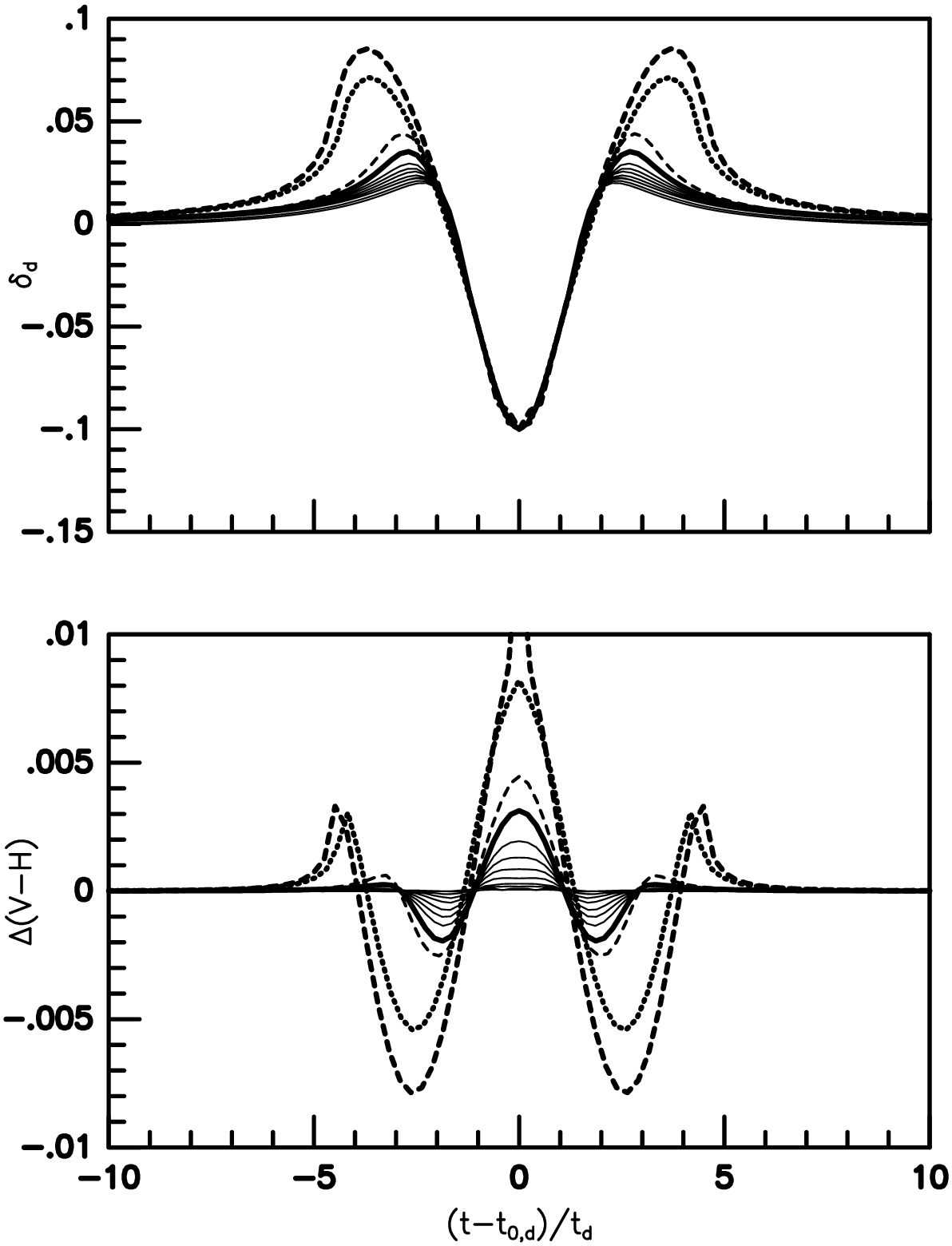}{6.4}{7.5}{6.5}
{
a) Ten light curves with $\gamma_-=1.67$, $\phi=90^\circ$, 
all with maximum 
deviation $\delta_d=-10\%$ and FWHM
$t_d = 0.06\,t_e$.  The ratios of source radius to planet Einstein ring range
from $\rho=0.1$ to $\rho=1.83$, the largest source radius consistent
with this maximum deviation.  The corresponding relative
values of $q=m/M$, and the
relative proper motion, $\mu/\mu_0$, are given in Table 4.
Source radii of $\rho = 1.6, 1.7, 1.83,$ and $1.80$ correspond
to the bold, dashed, bold dotted, and bold dashed curves.   
b) The fractional color change $\Delta(V-H)$ for the ten curves in panel (a).
}
\endinsert
 \topinsert
\mongofigure{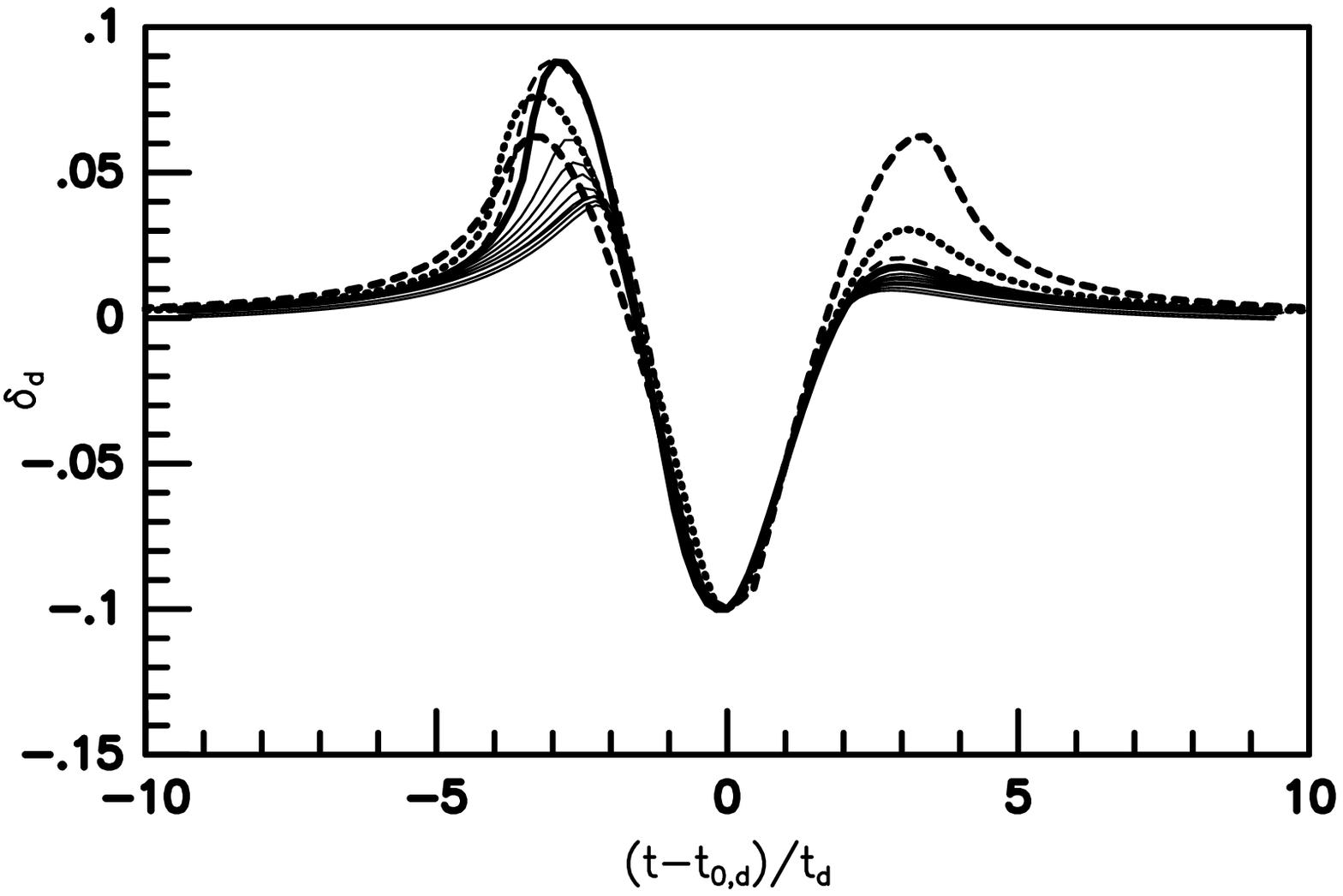}{6.4}{4.5}{6.5}
{
Ten light curves with $\gamma_-=0.6^{-1}$, $\phi=60^\circ$, 
all with maximum deviation $\delta_d=-10\%$ and FWHM
$t_d = 0.06\, t_e$.  All other parameters are the same
as Fig. \ten, and are given in Table 4.  Curves are
as in Fig. \ten.
}
\endinsert
Figure \ten a shows twelve light curves for $\gamma = 1.67$ and
$\phi = 90^\circ$, all with
maximum negative perturbation $\delta_d = -10\%$, and all with the
same FWHM.  Table 4 gives the inferred values of $q$ and $\mu$ for
each curve, relative to the fiducial 
values $q_0$ and $\mu_0$ associated with $\rho=0.3$. 
Note that the degeneracy in the derived mass ratios is only a 
factor of $\sim 1.5$.  Also note that the inferred mass ratios of 
the first nine curves agree to $\sim 4 \%$.  Thus to resolve the small
degeneracy in $q$, one only needs to distinguish between the
last four curves.  From Figure \ten a it is clear that this would be possible 
if one could resolve the positive perturbation structures at the $\sim 1\%$ level.
Furthermore, the situation presented in Figure \ten, for which $\phi = 90^\circ$,
is the worst case scenario.  Due to the structure of the caustics of
minor-image perturbations, trajectories with $\phi < 90^\circ$ display
marked asymmetry about $t_{0,d}$, excepting trajectories with $\alpha \sim 0$, which are 
nearly symmetric.  This enables one to distinguish between curves 
with $\alpha \gsim 1$ and $\alpha \sim 0$ more easily when $\phi < 90^\circ$.  This is demonstrated in
Figure \ell, which shows twelve light curves with the same parameters as
Figure \ten, except that now $\phi = 60^\circ$.  Comparing Figs.\ \ten\ and \ell,
it is clear that the curves are appreciably less degenerate for 
$\phi=60^\circ$ than for $\phi=90^\circ$.  From Figure \ten b, we see
that the magnitude of the fractional color change
for perturbations with $\gamma=1.67$ is always small, $\Delta(V-H) \lsim 1\%$.
From numerical calculations, we find that $\Delta(V-H) \lsim 1\%$ regardless
of the value of $\gamma$.  Thus, 
in contrast to major image perturbations,
optical/infrared color are not useful in resolving the degeneracy in minor image
perturbations, since the magnitude of the fractional color change is always small.

$$\vbox{\halign{#\hfil\quad&\hfil#\quad&\hfil#
\quad&\hfil#\hfil\quad&\hfil#\hfil\quad&\hfil#\hfil\cr
\multispan{4}{\hfil TABLE 4 \hfil}\cr
\noalign{\medskip}
\multispan{4}{\hfil Degenerate Parameter Values: 
Continuous Minor Image\hfil}\cr
\noalign{\smallskip}
\noalign{\hrule}
\noalign{\smallskip}
\noalign{\hrule}
\noalign{\smallskip}
\hfil impact \hfil&\hfil dimensionless \hfil&\hfil 
planet/star \hfil&\hfil proper motion \hfil\cr
\hfil parameter\hfil&\hfil source radius\hfil&\hfil 
mass ratio\hfil&\hfil \hfil\cr
\hfil $\alpha$\hfil&\hfil $\rho$ \hfil&\hfil $q/q_0$ \hfil&\hfil $\mu/\mu_0$ \hfil\cr
\noalign{\smallskip}
\noalign{\hrule}
\noalign{\smallskip}
\hfil4.32\hfil&\hfil0.10\hfil&\hfil0.993\hfil&\hfil4.014\hfil \cr
\hfil4.30\hfil&\hfil0.20\hfil&\hfil1.016\hfil&\hfil1.984\hfil \cr
\hfil4.25\hfil&\hfil0.40\hfil&\hfil1.000\hfil&\hfil1.000\hfil \cr
\hfil4.21\hfil&\hfil0.60\hfil&\hfil1.006\hfil&\hfil0.665\hfil \cr
\hfil4.00\hfil&\hfil0.80\hfil&\hfil1.020\hfil&\hfil0.495\hfil \cr
\hfil3.70\hfil&\hfil1.00\hfil&\hfil1.033\hfil&\hfil0.394\hfil \cr
\hfil3.50\hfil&\hfil1.20\hfil&\hfil1.017\hfil&\hfil0.331\hfil \cr
\hfil3.20\hfil&\hfil1.40\hfil&\hfil0.994\hfil&\hfil0.287\hfil \cr
\hfil2.60\hfil&\hfil1.60\hfil&\hfil1.004\hfil&\hfil0.249\hfil \cr
\hfil2.00\hfil&\hfil1.70\hfil&\hfil1.093\hfil&\hfil0.225\hfil \cr
\hfil1.00\hfil&\hfil1.83\hfil&\hfil1.591\hfil&\hfil0.173\hfil \cr
\hfil0.00\hfil&\hfil1.80\hfil&\hfil1.628\hfil&\hfil0.174\hfil \cr
\noalign{\smallskip}
\noalign{\hrule}
}}
$$

\chapter{From Mass Ratios to Planet Masses}

	If the various degeneracies described in this paper are broken, one 
generally recovers two planetary-system parameters from
a planetary microlensing event: $q$ and $y_p$.  While $q$ is of some interest
in its own right, $y_p$ is not.  The quantities one would most like to know
are the planet mass $m=q M$ and the physical projected separation 
$a_p = r_e y_p$.  One could take a purely statistical approach to estimating 
these quantities: given the measured time scale $t_e$ of the event and a 
plausible model of the distribution and velocities of lenses and sources
along the line of sight, $r_e$ and $M$ can be estimated to a factor of 3.
In this section, we discuss what further constraints might be
obtained on $M$ and $r_e$ in order to determine $m$ and $a_p$.

	The single most powerful method of acquiring additional information
would be to launch a parallax satellite
(Refsdal 1966; Gould 1995a; Gaudi \& Gould 1997) which would routinely measure
$\tilde r_e\equiv (\dos/\dls)r_e$ and often measure the direction of motion
as well.  This information would, by itself, narrow the uncertainty in the
mass to a factor $\sim 1.7$ (see Han \& Gould 1995, especially figure 7).
However, if the proper motion $\mu$ were also measured, this would yield
a complete solution of the lensing geometry including both $M$ and $r_e$
(e.g.\ Gould 1996).  In general, one expects to measure $\mu$ only in
$\sim 20\%$ of giant events even with relatively aggressive observations
(Gould 1996).  However, for events with planetary perturbations, $\mu$
can be measured much more frequently.  Recall from \S\ 6 that
for giant sources, major image perturbations,
 and planetary masses $m\lsim 100\,M_\oplus$, the planet
usually resolves the source (if it is detected at all) and that in the
process of resolving the resulting degeneracy, one measures $\mu$.
Even when $\theta_*<\theta_p$, the source will sometimes cross a caustic
in which case $\mu$ can be measured.  Finally, for $\theta_*<\theta_p$ one
can obtain a lower limit $\mu>\mu_\min$ based on the lack of detection of
finite-source effects.  Since the mass is given by
$M = (c^2/4 G)\tilde r_e t_e\mu$, and since $\tilde r_e$ and $t_e$ are
measured, this gives a lower limit on the mass (Gould 1995b).

	However, it is much more difficult to resolve the finite source
degeneracy for minor-image perturbations.  Even though
(or rather, because) the measurement
of the mass ratio, $q$, is not seriously hampered by this degeneracy,
the proper motion $\mu$ is poorly determined. See \S\ 2.3. Thus, it would be
necessary to measure $\mu$ using other methods.  See \S\ 6.1.

	We now address several questions related to one of those methods:
direct imaging of the source and lensing star several
decades after the event.  For definiteness, we suppose that the measurement
is made after 20 years. The expected separation is $\sim 0.\hskip-2pt ''1$,
but could plausibly be $\sim 0.\hskip-2pt ''3$.  At Baade's Window, the
expected number of stars $M_I<10$ inside this radius is $\sim 0.5$
(Light et al.\ 1996).  Thus one would not be overwhelmed with candidates.
On the other hand, the great majority of lensing events are almost certainly
due to objects that are fainter than $M_I=10$ simply because one does not
come close to accounting for the observed events from the observed ($M_I<10$)
stars alone (Han 1996).  Thus, to positively identify a candidate star
as the lens, one needs additional information.  A parallax satellite could 
provide two pieces of corroborating data.  First, the measured $\tilde r_e$
together with the proper motion inferred from the candidate-source angular
separation would give a mass and distance to the lens (Gould 1995b).  One could
then predict an apparent magnitude and see if it agreed with that of the
candidate.  Second, if the parallax measurement gave the angle of motion,
one could check this against the direction of the source-candidate
separation vector.  In addition, the candidate's inferred proper motion must
satisfy the lower limit derived from lack of finite-source effects as
discussed above.  Finally, one could wait another decade or so to see if the
direction of the candidate's proper motion was indeed away from the source.

{\bf Acknowledgements}:  
This work was supported in part by grant AST 94-20746 from the NSF,
and in part by grant NAG5-2864 from NASA.

\APPENDIX{A}{Justification for the Chang-Refsdal Approximation}

	What errors are introduced by the Chang-Refsdal approximation?  
The unperturbed image structure
consists of two images separated by $>2\theta_e$.  The planet, with an
effective sphere of influence $\sim \theta_p\ll \theta_e$ can have a major
effect on it at most one of these.  For definiteness, say this is the major
image.  In the Chang-Refsdal approximation, the minor image is then treated as
being completely unaffected by the presence of the planet.  In fact,
the planet will change the shear at the minor image by
${\cal O}
(\theta_p/\theta_e)$ and therefore change the magnification by a similar
amount.  However, what is directly of interest for analyzing the planetary
perturbation is not the absolute difference in magnification with and without
the planet.  Rather, it is the {\it change} in this difference over the
lifetime of the planetary perturbation.  Hence, the net effect is
${\cal O}[(\theta_p/\theta_e)^2]$, i.e., of higher order
than the effects being analyzed.  
	
	We now turn to the errors in the Chang-Refsdal estimate of the 
magnification of the perturbed image.  In general, the perturbed image is
split by the planet into two or four images.  For each such image, $i$, the
shear due to the parent star is $\gamma_i$.  If this value were exactly
equal to $\gamma$, the shear at the position of the unperturbed  image, then
the Chang-Refsdal approximation would be exact.  Typically, 
$\Delta\gamma_i\equiv \gamma_i -\gamma$ is small, 
$\Delta\gamma_i/\gamma\sim {\cal O}(\theta_p/\theta_e)$,
so one expects that the errors induced by the approximation are small.

 	We focus first on perturbations of the major image.  Let 
$\Delta\theta$ be the separation between the planet and the unperturbed image
and define $\alpha=\Delta\theta/\theta_p$.  Consider first the case
$\alpha\gg 1$ which is important when $\gamma\gsim 0.5$ because the 
magnification contours then become significantly elongated (see Fig.\ \three). 
The image is then split into two images, one very close to planet and the
other very close to the unperturbed image.  For the image close to the planet,
the shear due to  the parent star may be  significantly misestimated, 
$\Delta\gamma_i/\gamma\sim \alpha\theta_p/\theta_e$.  However, for this image,
the total shear is dominated by the planet and is ${\cal O}(\alpha^2)$, 
so true fractional error is only $\sim \alpha^{-1}\theta_p/\theta_e$.  
Moreover, the magnification of this image is small, ${\cal O}(\alpha^{-4})$,
so the total error induced by the approximation is 
$\sim \alpha^{-5}\theta_p/\theta_e$ and is completely negligible.  The other
image is displaced by $\sim\alpha^{-1}\theta_p$ from the unperturbed image,
so $\Delta\gamma_i/\gamma\sim \alpha^{-1}\theta_p/\theta_e$ which induces
a similar small change in magnification.  Recall from Figure \three-c, that the
source trajectory is determined up to a two-fold degeneracy from the maximum 
magnification.  Since the sign of the image displacement is different 
for the two allowed solutions, the error in estimating the magnification
structure could result in two types of errors.  First, there
is an error in the planet star separation, but this is only  
$\sim\alpha^{-1}\theta_p$ and is therefore lower by $\alpha^{-1}$ than the
basic degeneracy indicated in Figure \three-c.  Second, there is an error
in the estimate of $q$ and, in fact a degeneracy because the error has
opposite sign for the two allowed solutions.  This could in principle be
significant because, within the Chang-Refsdal 
framework, the two allowed solutions indicated in Figure \three-c\ have 
identical values of $q$, and this effect is therefore the lowest order
degeneracy.  However, the mass ratio is estimated from the FWHM of the light
curve which is only a weak function of position along the elongated 
magnification contours.  Moreover, the misestimate of that position is small.
We therefore estimate a fractional mass degeneracy of 
$\Delta q/q\sim \alpha^{-2} q^{1/2}$.  

	For $\alpha\lsim 1$ and sources that are small compared to the caustic
structure (seen e.g., in Fig.\ \three), the situation is similar to that of
caustic-crossing binary-lens events.  The light curves are highly 
non-degenerate, and one determines not only $q$ and $x_p$, but also $\rho$.
From the standpoint of understanding degeneracies, the important case is
when the source is of order or larger than the caustic.  Here, there are 
roughly equally magnified images displace roughly by $\theta_p$ on either
side of the planet. Hence, the lowest order errors cancel and the next
order errors are $\sim {\cal O}[(\theta_p/\theta_e)^2]$, and can therefore
be ignored.

	There is one exception to this conclusion. In the argument given above,
we implicitly assumed that the planetary perturbation would
be significant only over an interval of source motion $\sim\theta_p$.  This
assumption fails when the perturbation structure is elongated 
$(\gamma\gsim 0.5$) {\it and} when the angle of source motion is low
($\sin\phi=\beta/x_d\ll 1$).  In this case, the local shear is no longer
well approximated by the shear at the center of the perturbation.  A proper
calculation would then require that the shear be recalculated at every point
along the source trajectory, holding
 the planet fixed.  This was the approach of
Gould \& Loeb (1992) and the resulting magnification for fixed planet position
can be seen in their Figure 3.  [In the present work, by contrast, what is
held fixed in constructing Figs.\ \three\ and \four\ is the observable:
the shear at mid-point of the perturbation.]\ \  As can be seen by comparing
Figure 3-c of Gould \& Loeb (1992) and Figure \three-c of the present work,
for Jupiter mass planets the difference in contours can be significant.
However, there are three points to note.  First, such events are rare 
both because the conditions ($\gamma\gsim 0.5$, $\beta\ll x_d$) together imply
$\beta\lsim 0.2$ and because the elongated contours are encountered 
``edge on'', so the cross section is only $\sim \theta_p/\theta_e$.  Second,
the effect is proportional to $q^{1/2}$ and so would not be significant for,
e.g., Earth-mass planets.  Third, the nature of the effect is to provide
information to break degeneracies in cases when the Chang-Refsdal approximation
would lead one to believe that there is no information.  In brief, in certain
rare cases, the Chang-Refsdal approximation leads one to underestimate the
amount of information available.

	For perturbations of the minor image, the two principle sources of
degeneracy are first, confusion of the two caustic peaks with each other and 
second, confusion of one of these peaks with a perturbation of the major
image.  Because these peaks are offset in the direction perpendicular to
the star-planet axis, the error in their location is 
${\cal O}[(\theta_p/\theta_e)^2]$ and hence of higher order than their 
separation. As in the case of the major image, there are certain rare
events with $\sin\phi\ll 1$ for which the Chang-Refsdal approximation makes
the degeneracy seem somewhat worse than it is.

\endpage

\Ref\alard{Alard, C.\ 1996, in Proc. IAU Symp.\ 173,
Astrophysical Applications of Gravitational Lensing, p.\ 214,
 (Eds.\ C.\ S.\ Kochanek, 
J.\ N.\ Hewitt), Kluwer Academic Publishers}
\Ref\albrow{Albrow M., et al.\ 1996, in Proc. IAU Symp.\ 173,
Astrophysical Applications of Gravitational Lensing, p.\ 227
 (Eds.\ C.\ S.\ Kochanek, 
J.\ N.\ Hewitt), Kluwer Academic Publishers}
\Ref\Alcock{Alcock, C., et al.\ 1996, ApJ, submitted}
\Ref\Ansari{Ansari, et al.\ 1996, A\&A, in press}
\Ref\bandr{Bennett, D., \& Rhie, H.\ 1996, ApJ, in press}
\Ref\bandf{Bolatto, \& Falco, E.\ 1994, ApJ, 436, 112}
\Ref\cr{Chang, K., \& Refsdal, S.\ 1979, Nature, 282, 561}
\Ref\gaudi{Gaudi, B.\ 1996, in preparation}
\Ref\gandg{Gaudi, B., \& Gould, A.\ 1997, ApJ, 477, 000}
\Ref\goulda{Gould, A.\ 1995a, ApJ, 441, L21}
\Ref\gouldb{Gould, A.\ 1995b, ApJ, 447, 491}
\Ref\gouldb{Gould, A.\ 1996, PASP, 108, 465}
\Ref\gauch{Gould, A., \& Gaucherel,  1996, ApJ, 477, 000}
\Ref\gandl{Gould, A., \& Loeb, A.\ 1992, ApJ, 396, 104}
\Ref\gw{Gould, A., \& Welch, D.\ 1996, ApJ, 464, 212}
\Ref\ls{Loeb, A., \& Sasselov, D.\ 1995, ApJ, 449, 33L}
\Ref\mbg{Manduca, A., Bell, R.\ A., \& Gustafsson, B.\ 1977, A\&A, 61,809}
\Ref\mandp{Mao, S., \& Paczy\'nski, B.\ 1991, ApJ, 374, 37}
\Ref\pac{Paczy\'nski, B.\ 1986, ApJ, 304, 1}
\Ref\pratt{Pratt et al.\ 1996, in Proc. IAU Symp.\ 173,
Astrophysical Applications of Gravitational Lensing, p.\ 221
 (Eds.\ C.\ S.\ Kochanek, 
J.\ N.\ Hewitt), Kluwer Academic Publishers}
\Ref\Refs{Refsdal, S.\ 1964, MNRAS, 128, 295}
\Ref\sef{Schneider, P., Ehlers, J., \& Falco, E.\ E.\ 1992, Gravitational 
Lenses (Berlin: Springer-Verlag)}
\Ref\udal{Udalski, A., et al.\ 1994, Acta Astronomica, 44, 165}
\Ref\witt{Witt, H.\ 1995, ApJ, 449, 42}

\refout
\endpage
\bye